\documentclass[12pt,english,article,amsmath,amssymb]{article}
\pdfoutput=1 

\usepackage{pdfpages} 

\usepackage{graphicx,graphics}
\usepackage{dcolumn}
\usepackage{amsmath,amssymb,amsfonts}
\usepackage{latexsym,verbatim}
\usepackage{bm,bbm}
\usepackage{color}
\usepackage{ulem}

\usepackage[T1]{fontenc}
\usepackage[latin1]{inputenc}
\usepackage{dcolumn}
\usepackage{color}
\usepackage{babel}

\usepackage[percent]{overpic}
\newcommand{\bA}{{\bm A}}

\newcommand{\bq}{{\bm q}}
\newcommand{\br}{{\bm r}}

\newcommand{\vD}{v_{\mathrm{D}}}

\newcommand{\Real}{\mathrm{Re}}
\newcommand{\Imag}{\mathrm{Im}}

\DeclareMathAlphabet\mathbfcal{OMS}{cmsy}{b}{n}
\begin{document}
\title{Modulated phases of graphene quantum Hall polariton fluids}
\author{Francesco M.D. Pellegrino$^1$\footnote{Corresponding author: francesco.pellegrino@sns.it}, Vittorio Giovannetti$^1$,\\ Allan H. MacDonald$^2$, and Marco Polini$^{3}$}
\date{}
\maketitle
\noindent$^1$ NEST, Scuola Normale Superiore and Istituto Nanoscienze-CNR, Piazza dei Cavalieri 7, I-56126 Pisa, Italy. \\
$^2$ Department of Physics, University of Texas at Austin, Austin, Texas 78712-1081, USA. \\
$^3$ Istituto Italiano di Tecnologia, Graphene Labs, Via Morego 30, I-16163 Genova, Italy.\\
\maketitle

{\bf There is growing experimental interest in coupling cavity photons to the cyclotron resonance excitations of 
electron liquids in high-mobility semiconductor quantum wells or graphene sheets. 
These media offer unique platforms to carry out fundamental studies of exciton-polariton condensation and cavity quantum 
electrodynamics in a regime in which electron-electron interactions are expected to play a pivotal role. 
Focusing on graphene, we present a theoretical study of the impact of electron-electron interactions on a quantum Hall polariton 
fluid, that is a fluid of magneto-excitons resonantly coupled to cavity photons. 
We show that electron-electron interactions are responsible for an instability of graphene integer quantum Hall polariton 
fluids towards a modulated phase. 
We demonstrate that this phase can be detected by measuring the collective excitation spectra, 
which soften at a characteristic wave vector 
of the order of the inverse magnetic length.}\\

Fluids of exciton polaritons, composite particles resulting from coupling between electron-hole pairs (excitons) in semiconductors and cavity photons,
have been intensively investigated over the past decade~\cite{deng_rmp_2010,carusotto_rmp_2013,byrnes_naturephys_2014}. 
Because of the light mass of these quasiparticles, exciton polariton fluids display 
macroscopic quantum effects at standard cryogenic temperatures~\cite{deng_rmp_2010,carusotto_rmp_2013,byrnes_naturephys_2014}, in stark contrast to ultracold atomic gases. Starting from the discovery of Bose-Einstein condensation of exciton polaritons in 2006~\cite{kasprzak_nature_2006}, these fluids have 
been the subject of a large number of interesting experimental studies exploring, among other phenomena, superfluidity~\cite{amo_nature_2009,amo_naturephys_2009}, hydrodynamic effects~\cite{amo_science_2011}, Dirac cones in honeycomb lattices~\cite{jacqmin_prl_2014}, and logic circuits with minimal dissipation~\cite{ballarini_naturecommun_2013}.

The isolation of graphene~\cite{geim_naturemater_2007}---a two-dimensional (2D) honeycomb crystal of carbon atoms---
and other 2D atomic crystals~\cite{kostya_pnas_2005} including transition metal dichalcogenides~\cite{wang_naturenano_2012,chhowalla_naturechem_2013} and black phosphorus~\cite{fengnian_naturephoton_2014}, provides us with an enormously rich and tunable platform 
to study light-matter interactions and excitonic effects in 2D semimetals and semiconductors. 
Light-matter interactions in graphene in particular have been extensively explored over the past decade with both fundamental and applied 
motivations~\cite{bonaccorso_naturephoton_2010,grigorenko_naturephoton_2012,koppens_naturenano_2014,roadmap,fengnian_naturephoton_2014}.
Recent experimental advances have made it possible to monolithically integrate graphene with optical 
microcavities~\cite{engel_naturecommun_2012,furchi_nanolett_2012}, paving the way for fundamental studies of 
cavity quantum electrodynamics at the nanometer scale with graphene as the active medium. 
Progress has also been made using an alternate approach 
applied previously to conventional parabolic-band 2D electron liquids in semiconductor 
quantum wells~\cite{scalari_science_2012} by coupling graphene
excitations to the photonic modes of a Terahertz (THz) metamaterial 
formed by an array of split-ring resonators~\cite{valmorra_nanolett_2013}.

When an external magnetic field is applied to a 2D electron liquid in a GaAs quantum well~\cite{deliberato_prb_2010} or a graphene sheet~\cite{chirolli_prl_2012,pellegrino_prb_2014}, and electron-electron interactions are ignored, transitions 
between states in full and empty Landau levels (LLs) are dispersionless, mimicking the case of atomic 
transitions in a gas.  The cyclotron resonance of a 2D quantum Hall fluid can be tuned to resonance with the photonic modes of a 
cavity or a THz metamaterial~\cite{scalari_science_2012}, thereby establishing the requirements for {\it cavity quantum Hall electrodynamics} (cQHED). 
Cavity photons have already been used to carry out spectroscopic investigations of 
fractional quantum Hall fluids~\cite{smolka_science_2014}.
cQHED phenomena present several important twists on ideas from 
ordinary atom-based cavity quantum electrodynamics because in this case 
interactions between medium excitations are strong and long-ranged.  
Furthermore the active medium can be engineered  in interesting ways,
for example by using, instead of a single 2D crystal, van der Waals 
heterostructures~\cite{novoselov_ps_2012,bonaccorso_matertoday_2012,geim_nature_2013} or vertical heterostructures 
which include both graphene sheets and ordinary semiconductor 
quantum wells~\cite{principi_prb_2012,gamucci_naturecommun_2014}.

In this Article we show that electron-electron (e-e) interactions play a major qualitative role in graphene-based cQHED. Before describing the technical details of our calculations, let us briefly summarize the logic of our approach. The complex many-particle system of electrons in a magnetic field, interacting between themselves and with cavity photons, is treated within two main approximations. We use a quasi-equilibrium approach based on a microscopic grand-canonical Hamiltonian and treat interactions at the mean-field level. We critically comment on these approximations after the Results section. Our approach is similar to that used in Refs.~\cite{kamide_prl_2010,byrnes_prl_2010,kamide_prb_2011}, except that simplifications associated with Landau level quantization allow more steps in the calculation to be performed analytically.

The problem of finding the most energetically favorable state of a graphene integer quantum Hall polariton fluid (QHPF) is approached in a variational manner, by exploiting a factorized many-particle wave function. The latter is written as a direct product of a photon coherent state and a Bardeen-Cooper-Schrieffer state of electron-hole pairs belonging to two adjacent LLs. We find that e-e interactions are responsible for an instability of the uniform exciton-polariton condensate state  towards a weakly-modulated condensed state, which can be probed experimentally by using light scattering. We therefore calculate the collective excitation spectrum of the graphene QHPF by employing the time-dependent Hartree-Fock approximation. We demonstrate that the tendency to modulation driven by e-e interactions reflects into the softening of a collective mode branch at a characteristic wave vector of the order of the inverse magnetic length.

{\bf Results}

{\bf Effective model.} We consider a graphene sheet in the presence of a strong perpendicular magnetic field ${\bf B} = B {\hat {\bf z}}$~\cite{castroneto_rmp_2009,goerbig_rmp_2011}. We work in the Landau gauge with vector potential $\bA_0 = - B y {\hat {\bf x}}$. The magnetic field quantizes the massless Dirac fermion (MDF) linear dispersion into a stack of LLs, $\varepsilon_{\lambda, n} = \lambda \hbar\omega_{\rm c} \sqrt{n}$, which are labeled by a band index $\lambda = \pm$ which distinguishes conduction and valence band states and an integer $n = 0,1,2,\dots$. 
Here $\omega_{\rm c} =  \sqrt{2} \vD / \ell_B$ is the MDF cyclotron frequency~\cite{castroneto_rmp_2009,goerbig_rmp_2011}, 
$v_{\rm D} \simeq c/300$ the Dirac band velocity ($c$ being the speed of light in vacuum), and $\ell_B= \sqrt{\hbar c/(eB)} \simeq 25~{\rm nm}/\sqrt{B[{\rm Tesla}]}$ is
the magnetic length. The spectrum is particle-hole symmetric, i.e.~$\varepsilon_{-, n} = -\varepsilon_{+, n}$ for each $n$. 
Each LL has macroscopic degeneracy ${\cal N} = N_{\rm f} S /(
2 \pi \ell^2_B) \equiv N_{\rm f} {\cal N}_\phi$, where $N_{\rm f} = 4$ is the spin-valley degeneracy and $S=L^2$ is the sample area. 

In this Article we address the case of integer filling factors, which we expect to be most accessible experimentally.  
Because of particle-hole symmetry, we can assume without loss of generality that the chemical potential lies in the 
conduction band between the $n = M$ and $n = M +1$ Landau levels. When the energy $\hbar \omega$ of cavity photons is 
nearly equal to the cyclotron transition energy 
$\Omega_M \equiv \varepsilon_{+, M+1} - \varepsilon_{+, M} = \hbar\omega_{\rm c}(\sqrt{M+1}-\sqrt{M})$, the full fermionic Hilbert space can be effectively 
reduced to the conduction-band doublet $M,M+1$.  

We introduce the following effective grand-canonical Hamiltonian:
\begin{equation}\label{eq:H_tot}
 {\cal H} = {\cal H}_{\rm ph}+ {\cal H}_{\rm mat} + {\cal H}_{\rm int}  -\mu_{\rm e} N_{\rm e}-\mu_{\rm X} N_{\rm X}~.
\end{equation}

The first term, ${\cal H}_{\rm ph}$, is the photon Hamiltonian, 
${\cal H}_{\rm ph}= \sum_{\bq, \nu} \hbar \omega_{\bq} a_{\bq, \nu}^\dag a_{\bq, \nu}$, 
where $a^{\dag}_{\bq, \nu}$ ($a_{\bq, \nu}$) creates (annihilates) a cavity photon with wave vector $\bq$, 
circular polarization $\nu = {\rm L}, {\rm R}$, and frequency
$\omega_{\bq}=\sqrt{\omega^2+c^2 q^2/\kappa_{\rm r}}$,  $\kappa_{\rm r}$ being 
the cavity dielectric constant and $c$ the speed of light in vacuum.

The second term in Eq.~(\ref{eq:H_tot}), ${\cal H}_{\rm mat}$, is the matter Hamiltonian, which describes the 2D MDF quantum Hall fluid, and contains a term due to e-e interactions. This Hamiltonian is carefully derived in the Supplementary Note 1. In brief, one starts from the full microscopic Hamiltonian of a 2D MDF quantum Hall fluid~\cite{goerbig_rmp_2011}, written in terms of electronic field operators $c_{\lambda, n, k, \xi}$. Here, $\lambda = \pm$ is a conduction/valence band index, $n$ is a LL index,  $k=-L/(2 \ell_B^2) + (2 \pi / L) j$ with $j=1,\ldots,{\cal N}_{\phi}$ is the eigenvalue of the 
$\hat{\bf x}$-direction magnetic translation operator, and $\xi$ is a fourfold index, which refers to valley ($K,K^\prime$) and spin ($\uparrow,\downarrow)$ indices. All the terms that involve field operators $c_{\lambda, n, k, \xi}$, $c^\dag_{\lambda, n, k, \xi}$ acting only on the conduction-band doublet $M,M+1$ are then treated in an exact fashion, while all other terms are treated at leading order in the e-e interaction strength~\cite{Giuliani_and_Vignale}. 

The third term, ${\cal H}_{\rm int}$, describes interactions between electrons and cavity photons, which we treat in the rotating wave approximation. 
This means that in deriving ${\cal H}_{\rm int}$ we retain only terms that conserve the sum of the number of photons and the number of matter excitations. Details can be found in the Supplementary Note~2. It is parametrized by the following light-matter coupling parameter
\begin{equation}\label{eq:lightmattercoupling}
g_{{\bf q}}=\hbar \omega_{\rm c} \sqrt{\frac{e^2}{2 \kappa_{\rm r} L_z \hbar \omega_{{\bf q}}}}~.
\end{equation}
In Eq.~(\ref{eq:lightmattercoupling}) $L_z \ll L$ is the length of the cavity in the ${\hat {\bf z}}$ direction 
($V= L_z L^2$ is the volume of the cavity). In what follows we consider a half-wavelength cavity, setting $\omega = \pi c/(L_z\sqrt{\kappa_{\rm r}})$. Consequently~\cite{pellegrino_prb_2014}, $g \equiv g_{\bf 0} = \hbar\omega_{\rm c} \sqrt{\alpha_{\rm QED}/(2\pi\sqrt{\kappa_{\rm r}})}$, where $\alpha_{\rm QED} = e^2/(\hbar c)\simeq 1/137$ is the QED fine-structure constant.

Finally, in Eq.~(\ref{eq:H_tot}) we have introduced two Lagrange multipliers, $\mu_{\rm e}$ and $\mu_{\rm X}$, to enforce 
conservation of the average number of electrons and excitations~\cite{su_prl_2014}. $N_{\rm e}$ is the electron number operator in the $M, M+1$ reduced Hilbert space, while $N_{\rm X} = N_{\rm ph} + N_{\rm ex}$ is the operator for the number of matter excitations (excitons). The value of the chemical potential $\mu_{\rm e}$ should be fixed to enforce $\langle \psi|N_{\rm e}|\psi\rangle = {\cal N}$. At zero temperature this condition is simply enforced in the variational wave function defined below.

\begin{figure}[t]
\centering
\includegraphics[width=0.8\columnwidth]{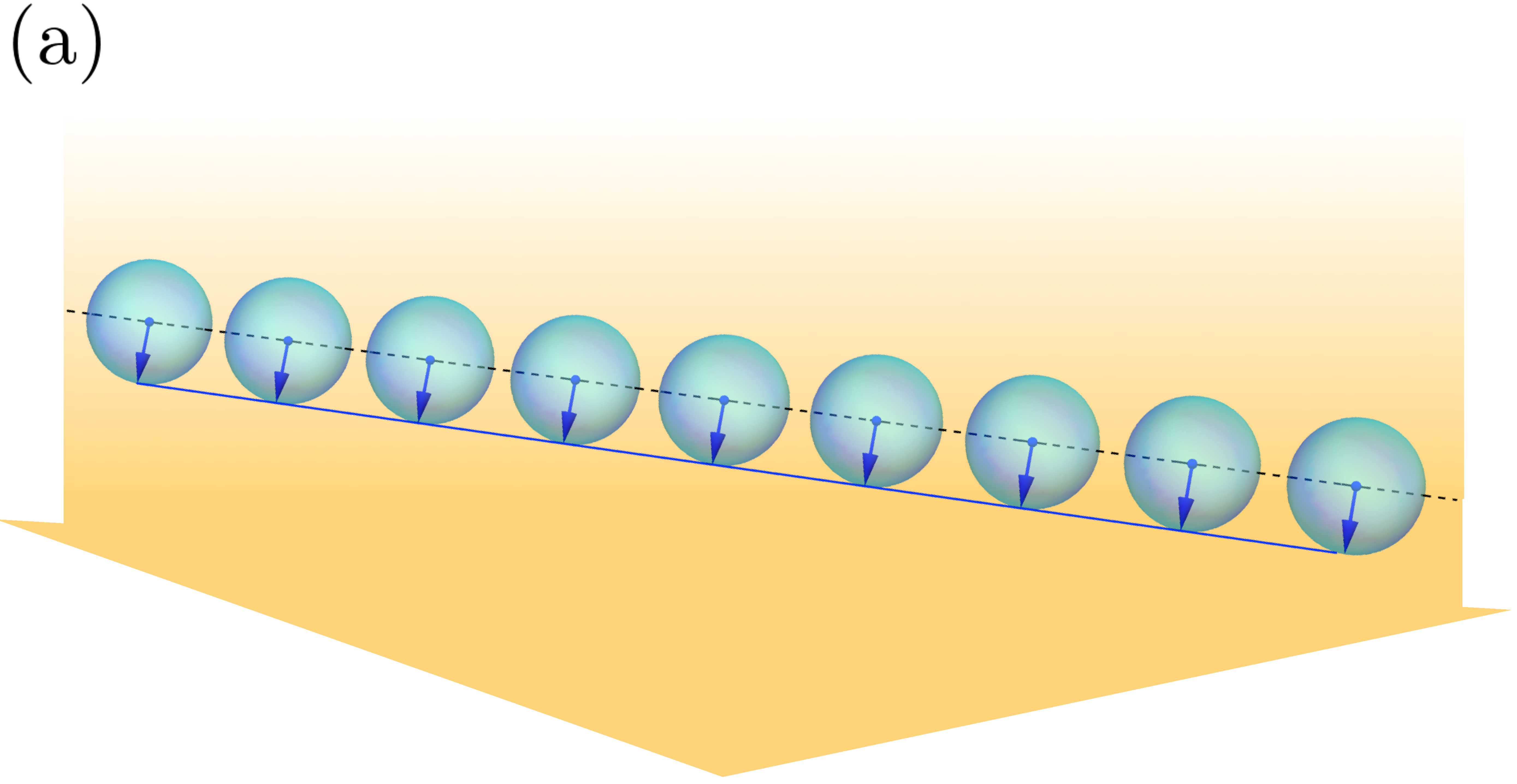}
\includegraphics[width=0.8\columnwidth]{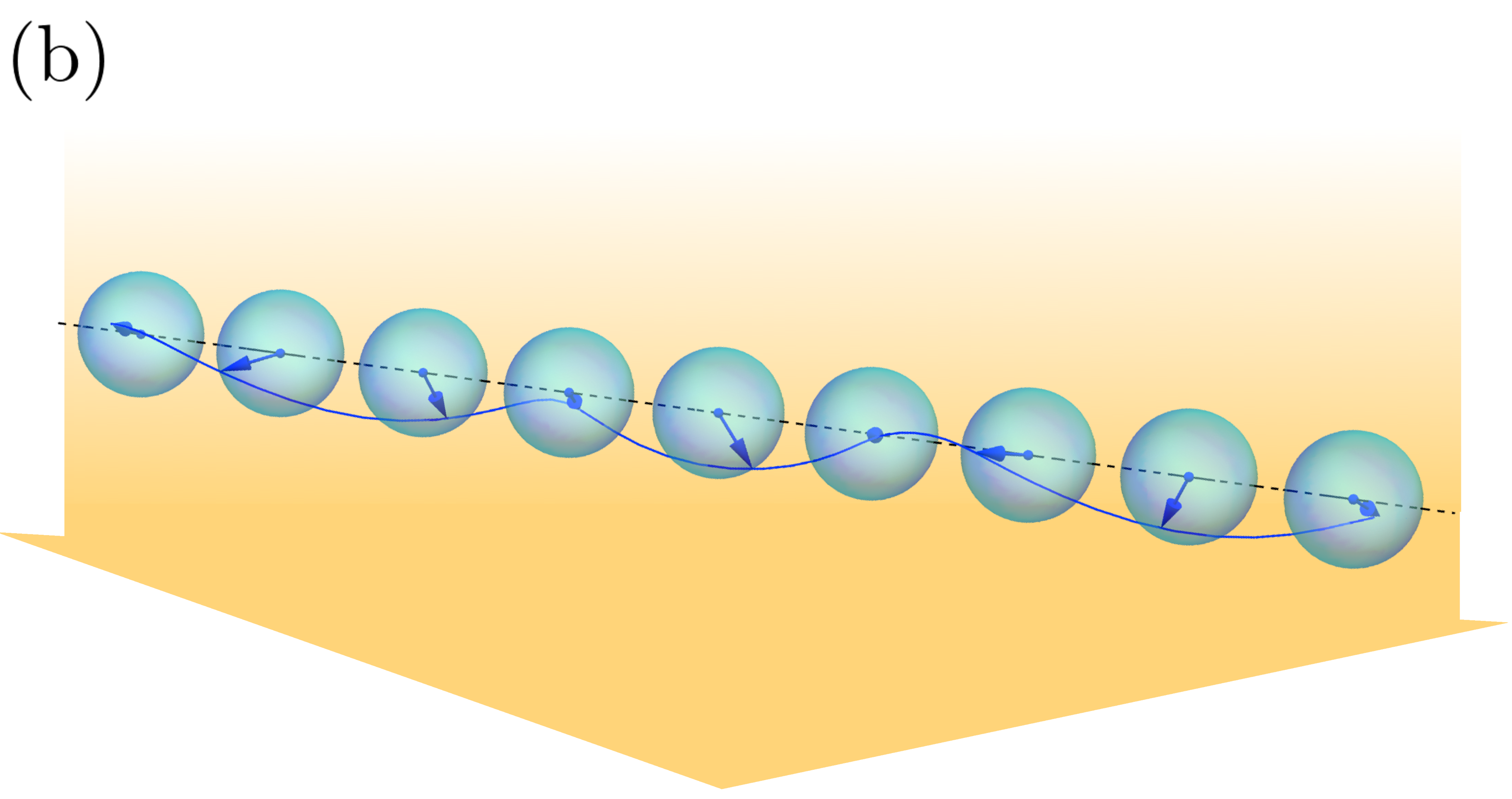}
\caption{{\bf Phases of a graphene integer quantum Hall polariton fluid.} 
Pictorial representation of the two phases supported by a graphene quantum Hall fluid interacting with a uniform electromagnetic field (in yellow). Panel ({\bf a}) When electron-electron interactions are weak, the ground state of the system $|\psi\rangle$ is a spatially-uniform polariton condensate. In pseudospin magnetic language, this state is a collinear ferromagnet with all pseudospins, defined in Eq.~(\ref{eq:spin}), denoted by blue arrows in the Bloch sphere, pointing along a common direction. Panel ({\bf b}) When electron-electron interactions are sufficiently strong, the ground state of the system $|\psi\rangle$ spontaneously break translational invariance. In
pseudospin magnetic language, this state is a spiral pseudospin state.  The spiral is driven by antisymmetric interactions 
between $\hat{\bf z}$ and $\hat{\bf x}$ pseudospin components as explained in the main text. \label{fig:cartoon}} 
\end{figure}

{\bf Variational wave function and spin-chain mapping.} 
To find the ground state of the Hamiltonian (\ref{eq:H_tot}), we employ a variational approach in which the many-particle wave function $|\psi \rangle$ is written 
as~\cite{eastham_prb_2001,byrnes_prl_2010} a direct product of a photon coherent state and a Bardeen-Cooper-Schrieffer state of electron-hole pairs belonging to the $M, M+1$ conduction-band doublet:
\begin{eqnarray}\label{eq:mfstate}
|\psi \rangle &=& \exp{\left(-{\cal N}\frac{|\alpha|^2}{2}\right)}\exp{\left(\sqrt{{\cal N}} \alpha~a^\dag_{{\bf 0}, {\rm L}}\right)}\nonumber\\
&\times&\prod_{k,\xi} \left[\cos(\theta_k/2)  + e^{-i \phi_k} \sin(\theta_k/2)c^\dag_{+, M+1, k, \xi} c_{+, M, k, \xi} \right] |\psi_0 \rangle~, 
\end{eqnarray}
where $ |\psi_0\rangle$ is the state with no photons and with the $M$-th LL fully occupied.
In writing Eq.~(\ref{eq:mfstate}), we have allowed for phase-coherent superposition of electron-hole pairs with $k$-dependent phases $\phi_k$ and excitation amplitudes $\sin(\theta_k/2)$, 
to allow for the emergence of modulated QHPF phases driven by e-e interactions.
Eq.~(\ref{eq:mfstate}) can be written in terms of polariton operators, as shown in Supplementary Note~3.
The variational parameters $\{\phi_k\}$, $\{\theta_k\}$, and $\alpha$ can be found by minimizing the ground-state energy $ \langle \psi |  {\cal H} | \psi \rangle$. 
We introduce the following regularized energy (per electron)~\cite{barlas_prl_2007}
\begin{equation}\label{eq:e_func}
 {\cal E} = {\cal E}(\{\phi_k\},\{\theta_k\},\alpha)\equiv \frac{\langle \psi|{\cal H}|\psi \rangle-\langle \psi_0|{\cal H}|\psi_0 \rangle}{{\cal N}}~.
\end{equation}
The variational wave function (\ref{eq:mfstate}) and the functional ${\cal E}(\{\phi_k\},\{\theta_k\},\alpha)$ can be conveniently expressed in terms of the $k$-dependent 
Bloch pseudospin orientations:
\begin{equation}\label{eq:spin}
{\bf n}(k) \equiv \frac{1}{N_{\rm f}}\sum_{\xi} \langle\psi|{\bf \sigma}_{k,\xi}|\psi\rangle =\left[\sin(\theta_k)\cos(\phi_k), \sin(\theta_k)\sin(\phi_k), - \cos(\theta_k)\right]^{\rm T}~,
\end{equation}
where ${\bf \sigma}_{k,\xi} \equiv \sum_{n, n^\prime = M, M+1}{\bf \tau}_{n n^\prime} c^\dag_{+, n,  k, \xi} c_{+, n^\prime,  k, \xi}$ and ${\bf \tau} = (\tau_1,\tau_2,\tau_3)$ is a 3D vector of Pauli matrices acting on the $M, M+1$ doublet. 
The variational wave function then becomes
\begin{eqnarray}\label{eq:rotspin}
|\psi \rangle &=& \exp{\left(-{\cal N}\frac{|\alpha|^2}{2}\right)}\exp{\left(\sqrt{{\cal N}} \alpha~a^\dag_{{\bf 0}, {\rm L}}\right)} \nonumber\\
&\times &
\exp\left(-i \sum_{k,\xi}  \theta_k {\bf m} (k) \cdot {\bf \sigma}_{k,\xi} /2 \right) | \psi_0 \rangle~,
\end{eqnarray}
where ${\bf m}(k) =\left(\sin(\phi_k), -\cos(\phi_k),0 \right)^{\rm T}$ is a unit vector orthogonal to ${\bf n}(k)$ and $| \psi_0 \rangle$ contains all pseudospins oriented along the $-\hat{\bf z}$ direction.
Since $\exp{[- i \theta_k  {\bf m} (k) \cdot {\bf \sigma}_{k,\xi}/2]}$ acts as a rotation by an angle $\theta_k$ around 
${\bf m}(k)$, we can interpret the matter part of $|\psi\rangle$ 
as a state in which every pseudospin labeled by $(k,\xi)$ is rotated accordingly. The unit vector ${\bf n}(k)$ in Eq.~(\ref{eq:spin}) denotes the final  pseudospin direction at each $k = 1\dots {\cal N}_\phi$. The string $\{{\bf n}(k)\}_k$ of ${\cal N}_\phi$ unit vectors 
can be viewed as a set of {\it classical} spins on a one-dimensional chain whose sites are labeled by the discrete index $k$, as in Fig.~\ref{fig:cartoon}. 

In the same notation,
\begin{equation} \label{eq:Ecal} 
{\cal E} = \epsilon - \epsilon_0 + \hbar \bar{\omega} n_{\rm X}~,
\end{equation}
where
\begin{eqnarray}\label{eq:eps}
\epsilon &\equiv&- g\frac{1}{{\cal N}_{\phi}} \sum_{k} {\bf B} \cdot {\bf n}(k) 
+ \frac{1}{2}\frac{1}{ {\cal N}_\phi^2} \sum_{k, k^\prime}\Bigg\{\sum^3_{\ell=1}  {\cal J}_{\ell}(k-k^\prime) n_{\ell}(k)n_{\ell}(k^\prime)\nonumber\\
&+&  \mathbfcal{D}(k-k^\prime)\cdot  \left[{\bf n}(k) \times {\bf n}(k^\prime)\right] \Bigg\}~,
\end{eqnarray}
$\hbar \bar{\omega} \equiv \hbar \omega_{{\bf q} = {\bf 0}} - \mu_{\rm X} = \hbar\omega - \mu_{\rm X}$, and
\begin{equation}\label{eq:constraint}
n_{\rm X} = \frac{\langle\psi|N_{\rm X}|\psi\rangle}{\cal N} = 
|\alpha|^2 + \frac{1}{{\cal N}_\phi} \sum_{k=1}^{{\cal N}_\phi} \frac{n_3(k)+1}{2}~. 
\end{equation}
In Eq.~(\ref{eq:eps}), the quantity $\hbar \bar{\omega}$ plays the role of a Lagrange multiplier, 
$\epsilon_0$ is a reference energy which is defined so that  that ${\cal E} = 0$ when $n_{\rm X}=0$, and $\mathbfcal{J} = ({\cal J}_1, {\cal J}_2, {\cal J}_3)$ and $\mathbfcal{D} = (0,{\cal D}_2,0)$ are the symmetric and antisymmetric, i.e. Dzyaloshinsky-Moriya (DM)~\cite{dzyaloshinsky_jpcs_1958,moriya_pr_1960}, interactions between Bloch pseudospins. 
Explicit expression for $\mathbfcal{D}$, $\mathbfcal{J}$, and  $\epsilon_0$ are provided in Supplementary Note~4, together with plots of the Fourier transforms ${\widetilde {\cal J}}_{\ell}(q) \equiv {\cal N}^{-1}_\phi \sum_k {\cal J}_{\ell}(k)\exp{(-i q k \ell^2_B)}$ and ${\widetilde {\cal D}_2}(q) \equiv {\cal N}^{-1}_\phi \sum_k {\cal D}_2(k)\exp{(-i q k \ell^2_B)}$
in Supplementary Figure~2.
In Eq.~(\ref{eq:constraint}) we note a photon contribution $n_{\rm ph} = \langle\psi|N_{\rm ph}|\psi\rangle / {\cal N} = |\alpha|^2$ and an exciton contribution, $n_{\rm ex}= \langle\psi|N_{\rm ex}|\psi\rangle / {\cal N} = {\cal N}_\phi^{-1}\sum_k \sin^2(\theta_k/2)$. 
It is somewhat surprising that DM interactions appear in our energy functional (\ref{eq:Ecal}), since these require spin-orbit interactions and appear when inversion symmetry is broken. Our microscopic Hamiltonian (\ref{eq:H_tot}) does not contain neither SOIs nor breaks inversion symmetry. In the next Section we discuss the origin of pseudospin DM interactions.

Each of the terms in the expression (\ref{eq:eps}) for $\epsilon = \epsilon(\{\phi_k\},\{\theta_k\},\alpha)$ has a clear physical interpretation. The first term on the right-hand side is the energy of a set of independent 1D Bloch pseudospins in an effective magnetic field with the usual Rabi coupling and 
detuning contributions: 
\begin{equation}\label{eq:extBfield}
{\bf B} \equiv \left[-\sqrt{2}\Real(\alpha),\sqrt{2} \Imag(\alpha),(\Delta-a_{\rm ee})/(2 g)\right]^{\rm T}~,
\end{equation}
where $\Delta \equiv \hbar \omega-(\Omega_M + \Delta_{\rm ee})$ is the detuning energy with 
$\Delta_{\rm ee}$ a correction due to e-e interactions between electrons
in the $M, M+1$ doublet and electrons in remote occupied LLs~\cite{iyengar_prb_2007,bychkov_prb_2008,shizuya_prb_2010} (see Supplementary Note~4). Because the MDF model applies over a large but {\it finite} energy interval, we need to introduce an ultraviolet cut-off $n_{\rm max}$ on
the LL labels $n$ of occupied states with $\lambda = -1$.  Our choice for $n_{\rm max}$ is explained in the Supplementary Note~5. 
It is easy to demonstrate that $\Delta_{\rm ee}$ depends logarithmically on 
$n_{\rm max}$: $\Delta_{\rm ee}= (\alpha_{\rm ee} \Omega_M/8) \left[ \ln(n_{\rm max})  + C_M \right]$
where $\alpha_{\rm ee}=e^2/(\kappa_{\rm r} \hbar \vD)$ is the graphene fine structure 
constant~\cite{kotov_rmp_2012} 
and $C_M$ is an ultraviolet-finite constant.  For $M=1$ we find that $C_1 \simeq -2.510$, in agreement with 
earlier work~\cite{shizuya_prb_2010}. 
The correction $\Delta_{\rm ee}$ to the cyclotron transition energy is related to the 
extensively studied~\cite{gonzalez_prbr_1999,borghi_ssc_2009,elias_naturephys_2011} renormalization of the Dirac 
velocity $\vD$ due to exchange interactions which also 
occurs in the absence of a magnetic field. 
The quantity $a_{\rm ee}$ involves only e-e interactions within the $M, M+1$ doublet (see Supplementary Note~4). 
For $M=1$ we find $a_{\rm ee}= \alpha_{\rm ee} \hbar \omega_{\rm c} (63/32+\sqrt{2})\sqrt{\pi}/32$.
The second term in Eq.~(\ref{eq:eps}) describes interactions between Bloch pseudospins,
which originate microscopically from matter-coherence dependence in the e-e interaction energy. 
At long wavelength these interactions stiffen the polariton condensate collective mode dispersion and support superfluidity. 
In the absence of a magnetic 
field their role at shorter wavelengths is masked by increasing exciton kinetic energy~\cite{wu_arxiv_soon}.

{\bf Pseudospin DM interactions.} 
In the QHPF exciton fluid kinetic energy is quenched and, as we explain below, DM exciton-exciton
interactions play an essential role in the physics. We therefore need to understand why $\mathbfcal{D}$ is finite. 
We start by observing (see Supplementary Note~4) that $\mathbfcal{D}$ 
contains direct ${\cal D}_{2, {\rm d}}$ and exchange ${\cal D}_{2, {\rm x}}$  contributions, which a) are of the same order of magnitude and b) have the same sign. We can therefore focus on the direct contribution, which has a simple physical 
interpretation as the electrostatic interaction between two charge distributions that are uniform along the 
 $\hat{\bf x}$ direction and vary along the $\hat{\bf y}$ direction, i.e.
\begin{equation}\label{eq:direct}
{\cal D}_{2, {\rm d}}(k-k^\prime)n_z(k)n_x(k^\prime) = - \frac{2}{\kappa_{\rm r} {\cal N} L}
\int d y d y^\prime \ln\left(\frac{|y-y^\prime|}{\ell_B}\right) \rho_z(y,k) \rho_x (y^\prime,k^\prime)
\end{equation}
where
\begin{eqnarray}\label{eq:rhoz}
\rho_z(y,k) &=&-e\frac{n_z(k)}{2}\Big\{\phi^2_{M+1}(y-k\ell^2_B) - \phi^2_{M}(y - k\ell^2_B) \nonumber \\
&+&  w^2_{-, M}\big[\phi^2_{M}(y-k\ell^2_B) - \phi^2_{M-1}(y- k\ell^2_B) \big] \Big\}
\end{eqnarray}
and
\begin{eqnarray}\label{eq:rhox}
\rho_x(y,k)&=&-en_x(k)\big[w_{+, M}\phi_{M+1}(y-k\ell^2_B) \phi_{M}(y-k\ell^2_B) \nonumber \\
&+&w_{-, M}\phi_{M}(y-k\ell^2_B) \phi_{M-1}(y-k\ell^2_B) \big]~.
\end{eqnarray}
Here $\phi_n(y)$ with $n= 0,1,2, \dots$ are normalized eigenfunctions of a one-dimensional harmonic oscillator with frequency $\omega_{\rm c}$ and $w_{\pm, n}=\sqrt{1 \pm \delta_{n, 0}}$ captures the property 
that the pseudospinor corresponding to the $n=0$ LL has weight only on one sublattice~\cite{goerbig_rmp_2011}. 

We now use a multipole expansion argument to explain  why ${\cal D}_{2, {\rm d}}(k-k^\prime)\neq 0$. We first note that $\rho_z(y,k)$ has zero electrical monopole and dipole moments but finite quadrupole moment 
$Q(k) \equiv -e  \ell^2_B Q_{M} n_z(k) = -e \ell_B^2 (1-\delta_{M,0}/2) n_z(k)$. On the other hand, $\rho_x(y,k)$ has zero 
electrical monopole but finite dipole moment 
$d(k) \equiv -e \ell_B d_M n_x(k) = -e \ell_B[w_{+, M}\sqrt{(M+1)/2} + \sqrt{M/2}] n_x(k)$. Using a multipole expansion, it 
follows that the leading contribution to Eq.~(\ref{eq:direct}) is the electrostatic interaction between a line of 
dipole moments extended along the $\hat{\bf x}$ direction and centered at one guiding center and 
and a line of quadrupole moments centered on the other guiding center. It follows that 
\begin{equation}
{\cal D}_{2, {\rm d}}(k-k^\prime) \approx- \frac{2 e^2}{\epsilon {\cal N} L \ell^3_B} \frac{Q_M d_M}{(k-k^\prime)^3}~. 
\end{equation}
The interactions are antisymmetric, {\it i.e.} their sign depends on whether the dipole is to the right or to the 
left of the quadrupole.  The direct contributions between like pseudospin components 
which contribute to $\mathbfcal{J}$ are symmetric 
because they are interactions between quadrupoles and quadrupoles or dipoles and dipoles.  

Alert readers will have noted that only the $\hat{\bf y}$-direction DM interaction is
non-zero, $\mathbfcal{D}(k-k^\prime)= {\cal D}_2(k-k^\prime) {\hat {\bf y}}$.
In contrast, the usual DM interaction~\cite{dzyaloshinsky_jpcs_1958,moriya_pr_1960} 
is invariant under simultaneous rotation of orbital and spin degrees of freedom. 
This is not the case for pseudospin DM interactions: the property that only the $\hat{\bf y}$ component of 
$\mathbfcal{D}$ is non-zero can be traced to the property that, for a given sign of pseudospin $n_x(k)$, the 
charge distribution $\rho_x(y, k)$ in Eq.~(\ref{eq:rhox}) changes sign under inversion around the 
guiding center (i.e.~$y \to -y +2 k \ell^2_B$).

{\bf Linear stability analysis of the uniform fluid state.} 
We first assume that the energy functional is minimized when $\theta_k$ and $\phi_k$ in Eq.~(\ref{eq:mfstate}) are $k$-{\it independent}, i.e.~$\theta_k = \theta$ and $\phi_k = \phi$ for every $k$. The functional ${\cal E}$ then simplifies to 
\begin{eqnarray}\label{eq:ener}
{\cal E}(\phi, \theta, \alpha)&=&  \hbar \bar{\omega}|\alpha|^2+  (\hbar \bar{\omega}- \Delta) \sin^2(\theta/2) + a_{\rm ee} \sin^4(\theta/2) \nonumber\\
&+& 2\sqrt{2} g|\alpha|  \cos(\phi+\arg(\alpha)) \sin(\theta/2)\cos(\theta/2)~.
\end{eqnarray}
The first term on the right hand side of Eq.~(\ref{eq:ener}), which is proportional to $|\alpha|^2$, is the free photon energy measured from the chemical potential $\mu_{\rm X}$. The second term, which is proportional to $\sin^2(\theta/2)$, is the free exciton energy (as renormalized by e-e interactions, which enter in the definition of $\Delta$). The third term, which is proportional to $\sin^4(\theta/2)$, is the exciton-exciton interaction term. Finally, the term in the second line, which is proportional to the Rabi coupling $\sqrt{2}g$, describes exciton-photon interactions.

We seek for a solution of the variational problem $\delta {\cal E} =0$ characterized by non-zero exciton and photon densities. For this to happen, 
the common chemical potential $\mu_{\rm X}$ needs to satisfy the following inequality:
\begin{equation}\label{eq:firstcondchempot}
|\hbar \bar{\omega}  ( \hbar \bar{\omega}- \Delta +a_{\rm ee})| < 2g^2 +\hbar \bar{\omega} a_{\rm ee}~.
\end{equation}
When this condition is satisfied, the solution of $\delta {\cal E} =0$ is given by
\begin{equation}\label{eq:argalpha}
\arg(\alpha) = \pi - \phi~,
\end{equation}
\begin{equation}\label{eq:modalpha}
|\alpha|^2=\frac{g^2}{2\hbar^2 \bar{\omega}^2}
\left\{1 -\left[\frac{\hbar \bar{\omega} (\hbar \bar{\omega}-\Delta + a_{\rm ee})}{2 g^2+\hbar \bar{\omega} a_{\rm ee}} \right]^2\right\}~,
\end{equation}
and
\begin{equation}\label{eq:sint}
\cos(\theta)=  \frac{\hbar \bar{\omega} (\hbar \bar{\omega}-\Delta + a_{\rm ee})}{(2 g^2+\hbar \bar{\omega} a_{\rm ee})}~.
\end{equation}
The common chemical potential $\mu_{\rm X}$ must be adjusted to satisfy $n_{\rm X}= [1 - \cos(\theta)]/2  + |\alpha|^2$. In the spin-chain language introduced above this state is a collinear ferromagnet in which all the classical spins $\{{\bf n}(k)\}_k$ are oriented along the same direction, as in Fig.~\ref{fig:cartoon}({\bf a}). Note that, as expected, the energy minimization problem does not determine the overall phase of the condensate.

We now carry out a local stability analysis to understand what is the region of parameter space in which this 
polariton state is a local energy minimum.  A minimum of ${\cal E}$, subject to the constraint on the average density $n_{\rm X}$ of excitations, 
is also a minimum of the functional $\epsilon(\{\phi_k\},\{\theta_k\},\alpha)$ defined in Eq.~(\ref{eq:eps}) with $|\alpha|$ not considered as an independent variable but rather viewed as a function of the variational parameters $\{\theta_k\}$ through the use of Eq.~(\ref{eq:constraint}), {\it i.e.} with 
~$|\alpha| \to |\alpha_{n_{\rm X}}(\{\theta_k\})| \equiv \sqrt{n_{\rm X} - {\cal N}^{-1}_{\phi} \sum_k \sin^2(\theta_k/2)}$. With this replacement,
\begin{equation}
{\widetilde \epsilon} \equiv \left.\epsilon(\{\phi_k\},\{\theta_k\}, \alpha)\right|_{|\alpha| \to |\alpha(\theta_k)|}
\end{equation}
becomes a functional of 
$2{\cal N}_\phi +1$ independent variational parameters, which can be arranged, for the sake of simplicity, into a vector ${\bf w}$ with components ${\bf w} =(\arg(\alpha), \theta_1,\ldots,\theta_{{\cal N}_\phi},  \phi_1,\ldots,\phi_{{\cal N}_\phi})^{\rm T}$. 

In this notation, the extremum discussed above 
can be represented by the vector ${\bf w}_0 =  (\pi-\phi, \theta,\ldots,\theta, \phi,\ldots,\phi)^{\rm T}$. 
We have checked that  ${\bf w}_0$ is a solution of the equation $\nabla_{\bf w} {\widetilde \epsilon}({\bf w})=0$.
Whether ${\bf w}_0$ is a local minimum or maximum depends on the spectrum of the Hessian
\begin{equation}\label{eq:hessian}
K_{mn}({\bf w}_0) = \left. \frac{\partial^2 {\widetilde \epsilon}({\bf w})}{\partial w_m \partial w_n}\right|_{{\bf w}= {\bf w}_0}~,
\end{equation}
which is a $(2{\cal N}_{\phi}+1) \times (2{\cal N}_{\phi}+1)$ symmetric matrix.

\begin{figure}[t]
\centering
\begin{minipage}{0.49\columnwidth}
\includegraphics[width=\columnwidth]{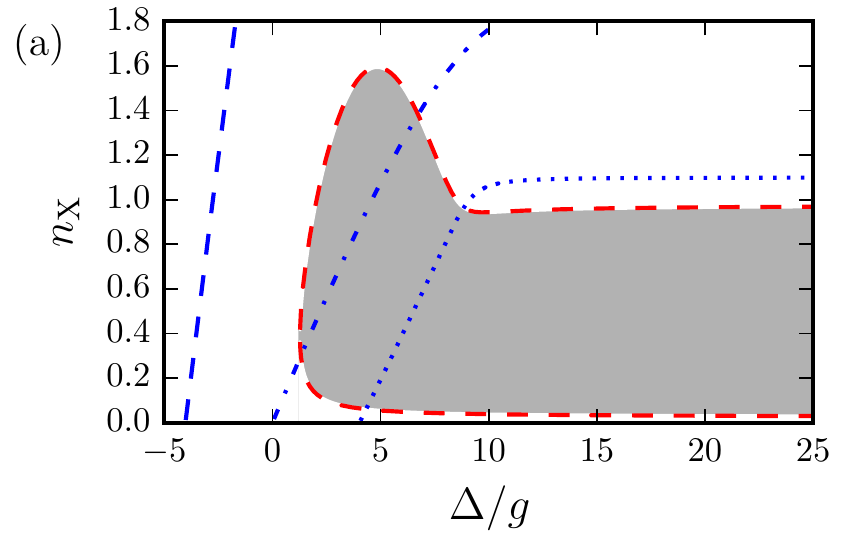}
\includegraphics[width=\columnwidth]{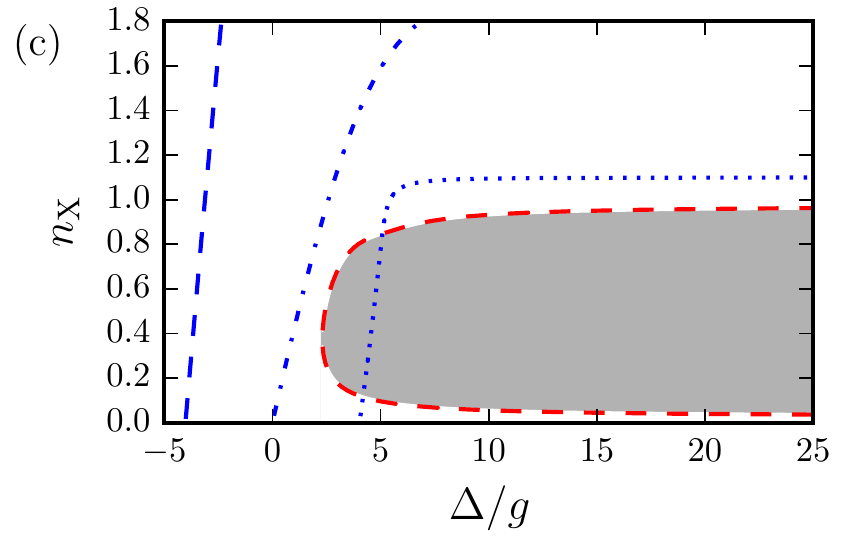}
\end{minipage}
\begin{minipage}{0.49\columnwidth}
\includegraphics[width=\columnwidth]{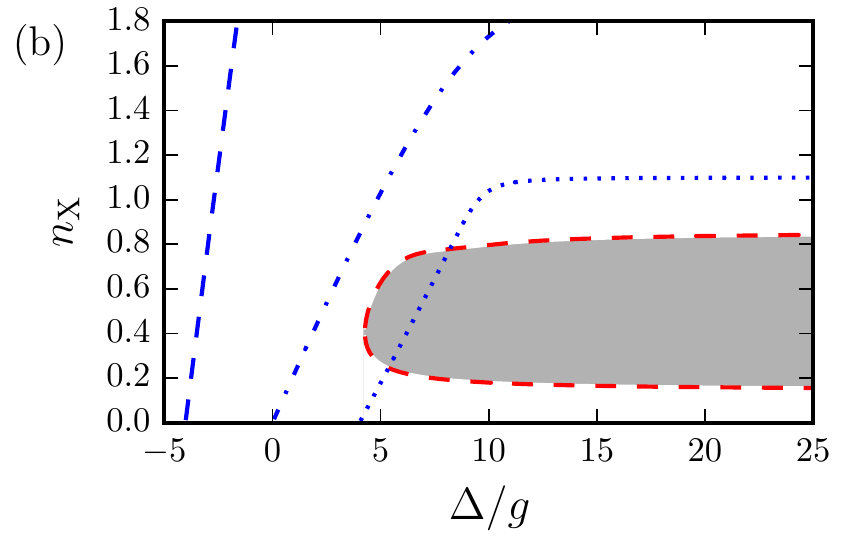}
\includegraphics[width=\columnwidth]{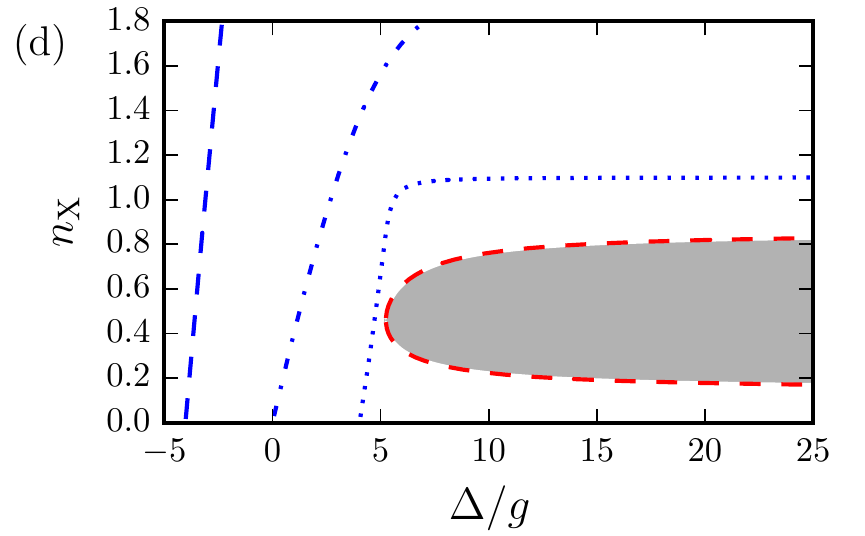}
\end{minipage}
\caption{{\bf Phase diagram of a graphene integer quantum Hall polariton fluid.} 
White (grey-shaded) regions represent the values of the detuning $\Delta$---in units of $g$---and density $n_{\rm X}$ of total excitations at which the homogeneous phase described by Eqs.~(\ref{eq:argalpha})-(\ref{eq:sint}) is stable (unstable). Panel ({\bf a}) $\kappa_{\rm r}=5$ and $M=1$; panel ({\bf b}) $\kappa_{\rm r}=5$ and $M=2$; panel ({\bf c}) $\kappa_{\rm r}=15$ and $M=1$; panel ({\bf d}) $\kappa_{\rm r} = 15$ and $M=2$.
In each panel, blue lines denote the location of points in the plane $(\Delta/g,n_{\rm X})$ where the ratio between the number of excitons and the number of photons is constant: $n_{\rm ex}/n_{\rm ph}=1/10$ (dashed line), $n_{\rm ex}/n_{\rm ph}=1$ (dash-dotted line), and $n_{\rm ex}/n_{\rm ph}=10$ (dotted line). These curves have been calculated with reference to the homogenous phase. \label{fig:cline_kr}}
\end{figure}

The homogeneous polariton fluid phase is stable only if $K_{mn}({\bf w}_0)$ has no negative eigenvalues. 
The stability analysis is simplified by exploiting translational symmetry to classify state fluctuations by momentum.  
Stability phase diagrams for $M=1$ and $M=2$, constructed by applying this criterion, 
are plotted in Fig.~\ref{fig:cline_kr} for two different values of the cavity dielectric constant $\kappa_{\rm r}$. 
In this figure, white (grey-shaded) regions represent the values of the detuning 
$\Delta$ and density $n_{\rm X}$ of total excitations for which the homogeneous fluid phase is stable (unstable).
As expected, by increasing $\kappa_{\rm r}$ (i.e.~reducing the importance of e-e interactions) the stable regions expand at the expense of the unstable ones. Note that the instability displays an intriguing {\it reentrant} character and that it can occur also when matter and light have comparable weigth, i.e.~when $n_{\rm ex} \sim n_{\rm ph}$.
 
We have checked that the root of instability of the homogeneous fluid phase is e-e interactions. More precisely, it is possible to see that in the absence of DM interactions---{\it i.e.} when ~$\mathbfcal{D}={\bf 0}$ in Eq.~(\ref{eq:eps})---the instability disappears. 
Symmetric interactions, however, still play an important quantitative role in the phase diagrams, as explained in Supplementary Note~6.
The physics of these phase diagrams is discussed further below where we identify the phase diagram boundary with the 
appearance of soft-modes in the uniform polariton fluid collective mode spectrum.  
Stable phases occur only if $\bar{\omega} > 0$ (i.e.~$\mu_{\rm X} < \hbar \omega$).  
We remind the reader that this condition on $\mu_{\rm X}$ is additional to the one given in Eq.~(\ref{eq:firstcondchempot}) above.

{\bf Elementary excitations of the polariton fluid.} 
We evaluate the elementary excitations of the uniform polariton fluid 
~\cite{deng_rmp_2010,carusotto_rmp_2013,byrnes_naturephys_2014} by linearizing the Heisenberg coupled 
equations of motion of the matter Bloch pseudospin and photon operators using a Hartree-Fock factorization for the 
e-e interaction term in the Hamiltonian (see Supplementary Note~7).  The collective excitation energies diagonalize the matrix  
\begin{equation}\label{eq:collective}
{\bf M} = \left(
 \begin{matrix}
\hbar \bar{\omega}_{{\bf q}} & 0 & - g^{\ast}_{x^\prime, {\bf q}} & -g^{\ast}_{y^\prime , {\bf q}}\\
0 & -\hbar \bar{\omega}_{{\bf q}} & g_{x^\prime , {\bf q}} &  g_{y^\prime , {\bf q}}\\
2 i g_{y^\prime , {\bf q}} & 2i g^{\ast}_{y^\prime , {\bf q}} & - i2 J_{x^\prime y^\prime}({\bf q})& - i[\Omega_{\rm MF} + 2J_{y^\prime y^\prime}({\bf q})]\\
- 2ig_{x^\prime, {\bf q}} & - 2i g^{\ast}_{x^\prime, {\bf q}}& i[\Omega_{\rm MF} + 2J_{x^\prime x^\prime}({\bf q})]& i2 J_{y^\prime x^\prime}({\bf q})
\end{matrix}
\right)~.
\end{equation}
The first two-components of eigenvectors of $\bf{M}$ correspond to photon creation and annihilation, and the third and fourth to rotations of the Bloch pseudospin in a plane (denoted by ${\hat {\bf x}}^\prime$-${\hat {\bf y}}^\prime$ in the Supplementary Note~7) orthogonal to its ground state orientation.  
In Eq.~(\ref{eq:collective}) $\hbar {\bar \omega}_{\bf q} = \hbar\omega_{\bf q} - \mu_{\rm X}$, 
$g_{x^\prime, \bq} \equiv e^{i \phi} \cos (\theta) g_\bq/\sqrt{2}$, $g_{y^\prime, \bq} \equiv i e^{i \phi} g_\bq/\sqrt{2}$,
\begin{equation}
\Omega_{\rm MF} = \frac{2g^2}{\hbar \bar{\omega}} - 2\widetilde{\cal J}_1(0)~,
\end{equation}
\begin{eqnarray}
J_{x^\prime x^\prime}({\bf q}) &=& [\widetilde{{\cal J}}_1(q)\sin^2(\varphi_{\bf q} - \phi)+\widetilde{{\cal J}}_2(q)\cos^2(\varphi_{\bf q} - \phi)]\cos^2( \theta) \nonumber\\
&+& \widetilde{\cal J}_3 (q) \sin^2( \theta)~,
\end{eqnarray}
\begin{equation}
J_{y^\prime y^\prime}({\bf q}) = \widetilde{{\cal J}}_2(q)\sin^2(\varphi_{\bf q} - \phi)+\widetilde{{\cal J}}_1(q)\cos^2(\varphi_{\bf q} - \phi)~,
\end{equation}
\begin{equation} 
\Re e[J_{x^\prime y^\prime}({\bf q})] = \sin(2\varphi_{\bf q} - 2\phi)\cos(\theta) [\widetilde{\cal J}_2 (q)-\widetilde{\cal J}_1 (q)]/2~,
\end{equation}
\begin{equation}\label{eq:imjxprimeyprime}
\Im m[J_{x^\prime y^\prime}(\bq)] = \widetilde{\cal D}(q) \cos(\varphi_\bq - \phi) \sin ( \theta)~,
 \end{equation}
and $\bq=[q\cos(\varphi_\bq),q\sin(\varphi_\bq)]^{\rm T}$. 

\begin{figure}[t]
\centering
\begin{minipage}{0.8\columnwidth}
\includegraphics[width=\columnwidth]{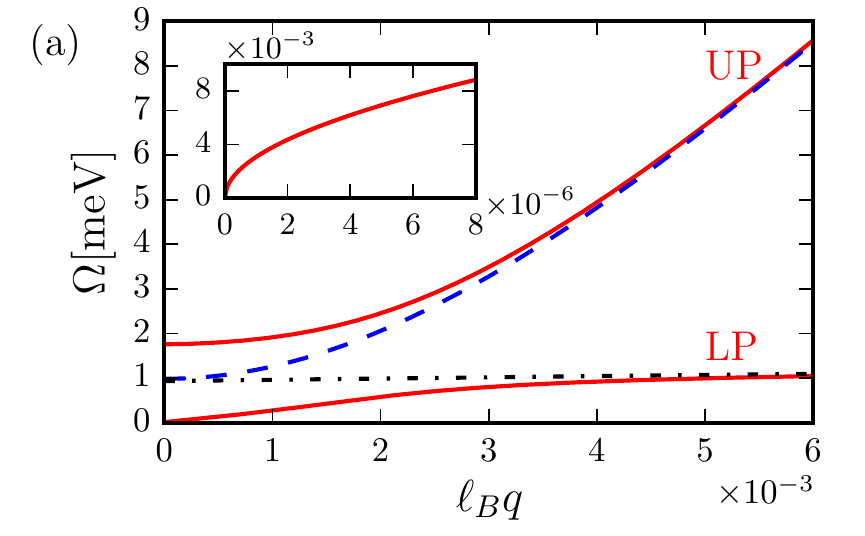}
\end{minipage}
\begin{minipage}{0.8\columnwidth}
\includegraphics[width=\columnwidth]{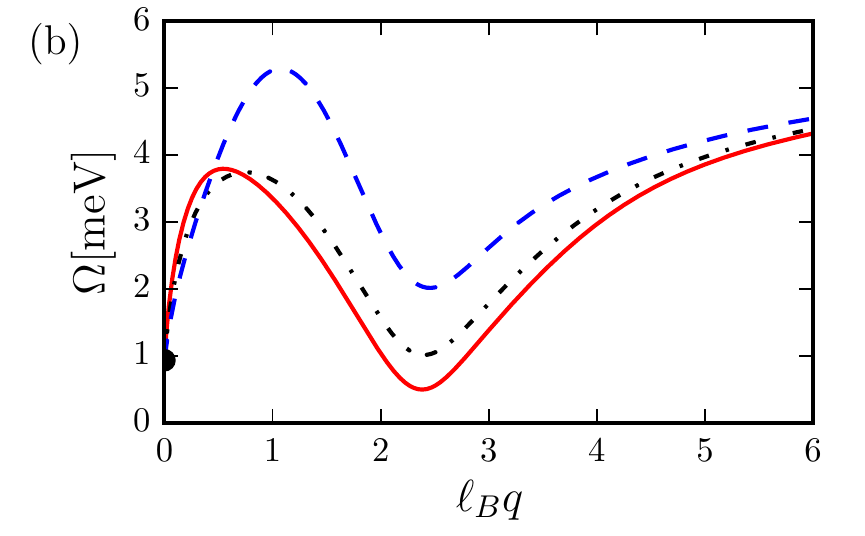}
\end{minipage}
\caption{{\bf Collective excitation spectrum of the homogeneous fluid phase.} Panel ({\bf a}) 
Dispersion relations of upper and lower dressed polariton modes (solid lines) in the long-wavelength $q \ell_B \ll 1$ limit.
The dashed line (dash-dotted line) represents the cavity photon dispersion (bright electronic collective mode), when the electron-photon coupling $g_{\bf q}$ is set to zero. In this panel $\varphi_{\bq} = \phi$. The inset shows a zoom of the lower polariton dispersion relation for $q\ell_{\rm B} \to 0$. Note that the dispersion behaves as $\sqrt{q}$ in this limit.
Panel ({\bf b}) Dispersion relation of the lower polariton mode, 
along three directions: $\varphi_{\bq}=\phi$ (solid line),
$\varphi_{\bq} = \phi+\pi/4$ (dash-dotted line), and $\varphi_{\bq} = \phi+\pi/2$ (dashed line). All the data in this figure have been obtained by setting $\kappa_{\rm r}=5$,  $B =0.5~{\rm Tesla}$, $M=1$, $n_{\rm X}=0.1$, and $\Delta = g$.\label{fig:modes}}
\end{figure}

The solution of the eigenvalue problem yields two hybrid modes that can be viewed as lower polaritons (LP) and upper polaritons (UP) that are dressed by the condensate and have strong mixing between photon and matter degrees of freedom at
$\ell_B q \lesssim v_{\rm D}\sqrt{ \kappa_{\rm r}}/c\ll 1$.
Fig.~\ref{fig:modes} illustrates the dispersion relations of these two modes for $M=1$. 
For wavelengths comparable to the magnetic length, $q \ell_B \sim 1$, the UP mode has nearly pure photonic character, while the LP mode is a nearly pure matter excitation with a dispersion relation that is familiar from the theory of magnon energies in systems with asymmetric DM exchange interactions~\cite{cote_prb_2010}:  
\begin{eqnarray}
\Omega_{{\bf q}} &\to& 2 \Im m[J_{x^\prime y^\prime}({\bf q})] \nonumber\\ 
&+& \sqrt{[\Omega_{\rm MF} + 2  J_{x^\prime x^\prime}({\bf q})][\Omega_{\rm MF} + 2  J_{y^\prime y^\prime}({\bf q})] - \{2 \Re e[J_{x^\prime y^\prime}({\bf q})]\}^2}~.
\end{eqnarray}
Fig.~\ref{fig:modes}({\bf b}) shows the LP dispersion relation for three different polar angles $\varphi_{\bq}$.
In all cases, a local roton-like minimum occurs at a wave vector $2/\ell_B< q <3/\ell_B$. 
The global minimum of the LP dispersion occurs along the direction $\varphi_{\bq}=\phi$, where the impact of DM interactions is strongest, i.e.~$\Im m[J_{x^\prime y^\prime}(\bq)]$ is maximum---see Eq.~(\ref{eq:imjxprimeyprime}).  The mode energy vanishes, and a Hessian eigenvalue crosses from positive 
to negative signalling instability, when
\begin{equation}\label{eq:softening}
 [\Omega_{\rm MF} + 2  J_{x^\prime x^\prime}({\bf q})][\Omega_{\rm MF} + 2  J_{y^\prime y^\prime}({\bf q})] \le  |J_{x^\prime y^\prime}({\bf q})|^2~.
\end{equation}
Since the LP mode becomes unstable at a {\it finite} wave vector $\sim 1/\ell_B$, we conclude that the true ground state 
spontaneously breaks translational symmetry.  We emphasize that softening of collective modes in quantum Hall fluids can be experimentally 
studied, for example, by inelastic light scattering~\cite{pellegrini_pss_2006}. 

{\bf Modulated phase of QHPFs.} Motivated by the properties of magnetic systems with strong asymmetric spin interactions~\cite{cote_prb_2010}, we seek broken translational states in which the Bloch pseudospins execute a small amplitude spiral around a mean orientation, as in Fig.~\ref{fig:cartoon}({\bf b}). This is a state in which $\theta_k,\phi_k$ have a rather simple $k$-dependence of the form: 
\begin{equation}\label{eq:Qmod}
\left\{
\begin{array}{l}
{\displaystyle \theta_k = \theta + u \cos(\ell^2_B Q^\ast k )}\vspace{0.2 cm}~,\\
{\displaystyle \phi_k   = \phi - v \sin(\ell^2_B Q^\ast k+ \varphi)}~.
\end{array}
\right.
\end{equation}
Eq.~(\ref{eq:Qmod}) physically describes a small-amplitude spatially-periodic contribution to the uniform condensate state (\ref{eq:mfstate}) with $\theta_k =\theta$ and $\phi_k=\phi$. One should therefore not confuse the condensed state described by Eqs.~(\ref{eq:mfstate}) and~(\ref{eq:Qmod}) with a uniform condensate in which electrons and holes form pairs with a finite center-of-mass momentum~\cite{casalbuoni_rmp_2004}.

Because the form factors of electrons in $M$ and $M+1$ LLs differ, this state 
has non-uniform electron charge density with periodicity $(2\pi/Q^\ast) \hat{\bf y}$. 
The Fourier transform of the density variation
\begin{equation}\label{eq:crystal}
\delta n_{\bf q} = \int d^2{\bf r}~e^{i {\bf q} \cdot {\bf r}}\left[\langle\psi|\Psi^\dag({\bf r})\Psi({\bf r})|\psi\rangle - \langle\psi_0|\Psi^\dag({\bf r})\Psi({\bf r})|\psi_0\rangle\right]~,
\end{equation}
is non-zero only for ${\bf q} = (0, n Q^\ast)$, where $n$ is a relative integer. 
In Eq.~(\ref{eq:crystal}) 
$\Psi({\bf r}) = \sum_{\lambda, n, k, \xi} \langle{\bf r}|\lambda,n,k\rangle c_{\lambda, n, k, \xi}$ with
\begin{equation}\label{eqn:LL_rspace}
\langle \br |\lambda, n, k \rangle= \frac{e^{i k x}}{\sqrt{2L}}\left(
 \begin{array}{c}
w_{-, n} \phi_{n-1}(y-\ell_B^2 k)\\
\lambda w_{+, n} \phi_{n}(y-\ell_B^2 k)
\end{array}
\right)~,
\end{equation}
is a field operator that creates an electron at position ${\bf r}$~\cite{goerbig_rmp_2011}. 

\begin{figure}[t]
\centering
\begin{minipage}{0.49\columnwidth}
\includegraphics[width=\columnwidth]{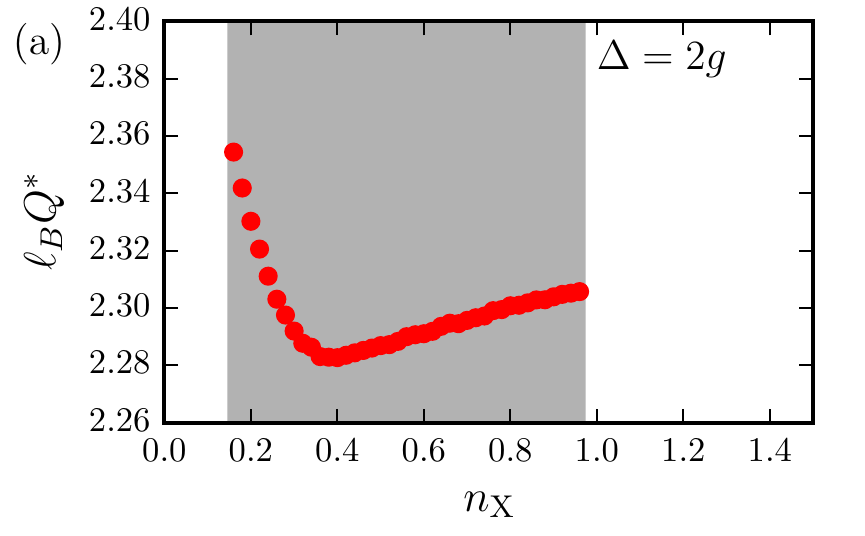}
\includegraphics[width=\columnwidth]{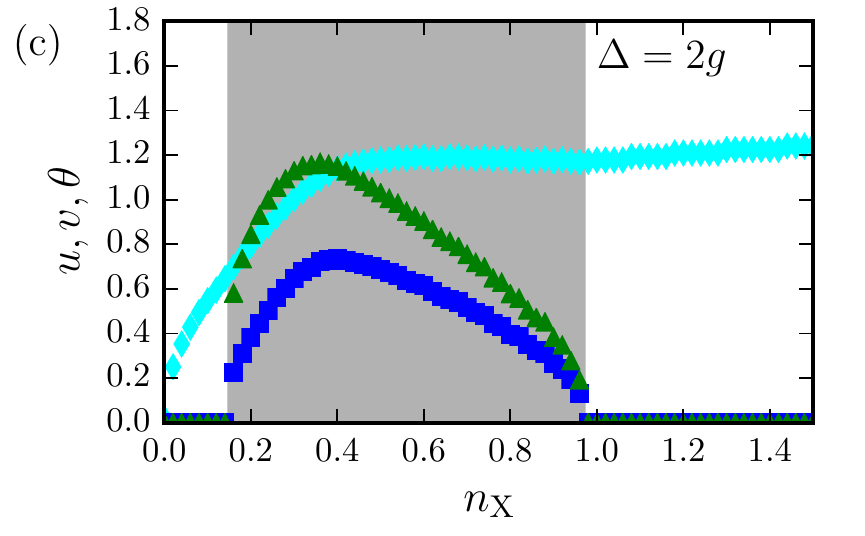}
\end{minipage}
\begin{minipage}{0.49\columnwidth}
\includegraphics[width=\columnwidth]{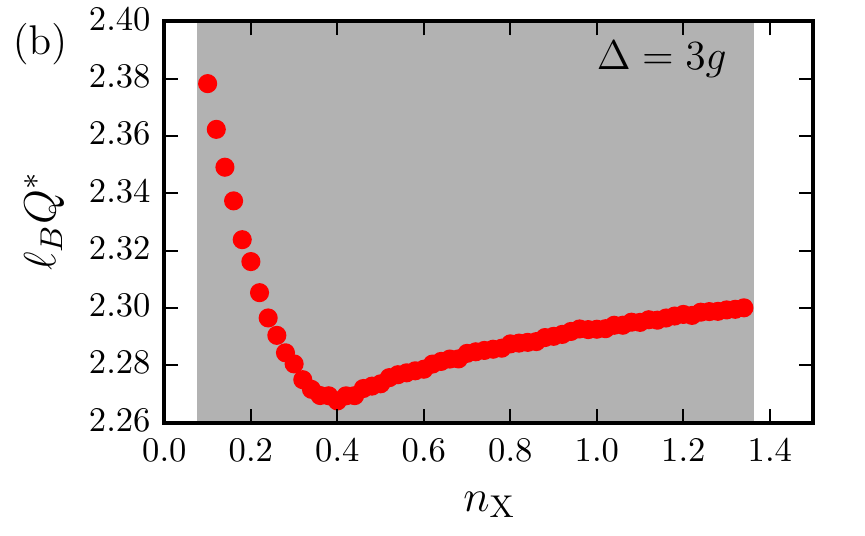}
\includegraphics[width=\columnwidth]{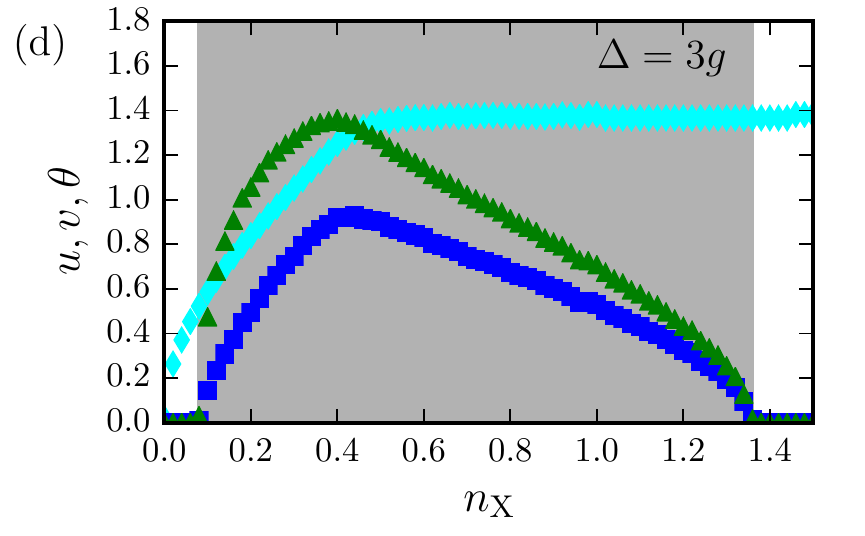}
\end{minipage}
\caption{{\bf Variational parameters of the modulated phase.} This figure shows the optimal values 
of the variational parameters $u,v,\theta$ and $Q^\ast$ in Eq.~(\ref{eq:Qmod}) for a cavity dielectric constant $\kappa_{\rm r}=5$,  highest occupied LL $M=1$ 
and different values of the detuning $\Delta$. Panels ({\bf a}) and ({\bf c}): $\Delta = 2g$. 
Panels ({\bf b}) and ({\bf d}): $\Delta = 3g$. Panels ({\bf a})-({\bf b}) Dependence of the characteristic wave number $Q^\ast$ (in units of $1/\ell_B$) on the density $n_{\rm X}$ of total excitations. Panels ({\bf c})-({\bf d}) Dependence of the quantities $u$ (blue squares), $v$ (green triangles), 
and $\theta$ (cyan diamonds) on the density $n_{\rm X}$ of total excitations. Grey-shaded areas have the same meaning as in Fig.~\ref{fig:cline_kr}.\label{fig:jn_kr5}}
\end{figure}

\begin{figure}[t]
\centering
\begin{minipage}{0.49\columnwidth}
\includegraphics[width=1.02\columnwidth]{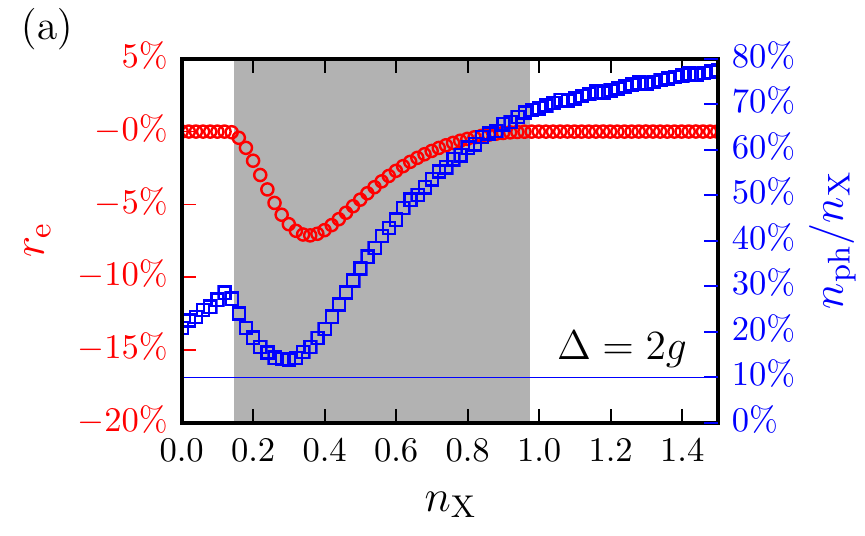}
\includegraphics[width=1.02\columnwidth]{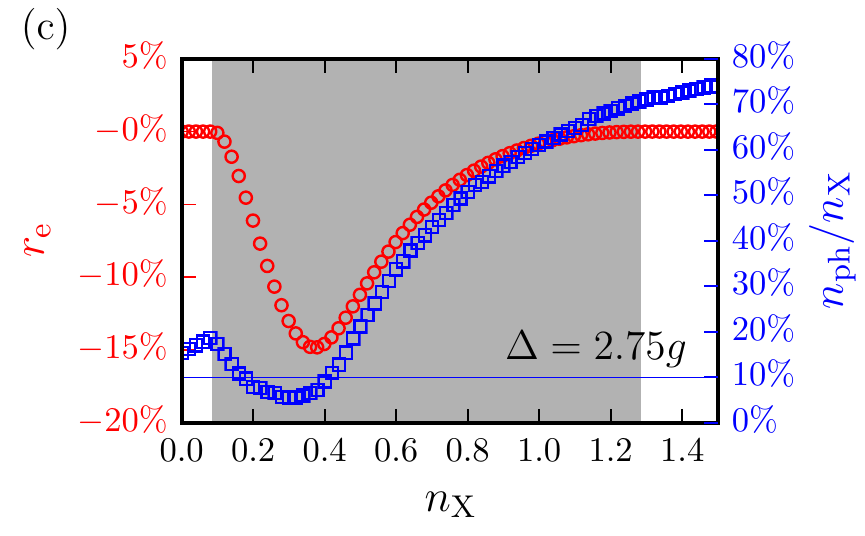}
\end{minipage}
\begin{minipage}{0.49\columnwidth}
\includegraphics[width=1.02\columnwidth]{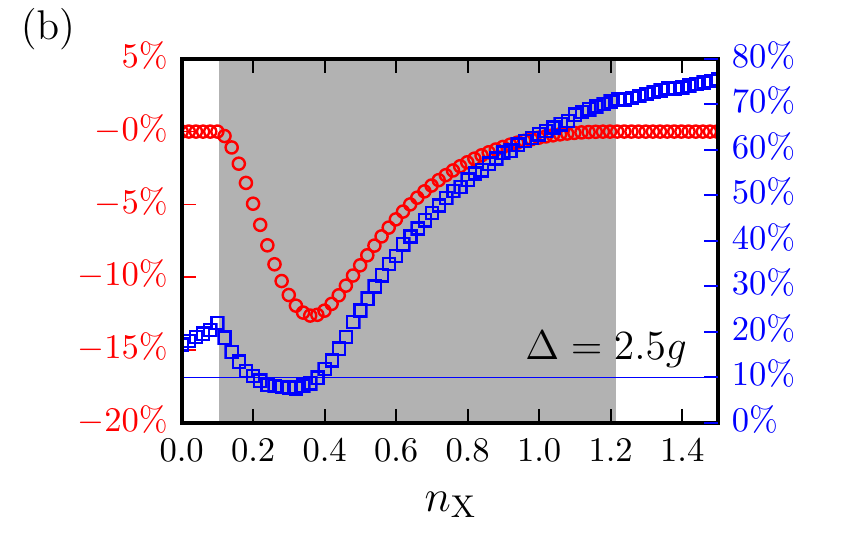}
\includegraphics[width=1.02\columnwidth]{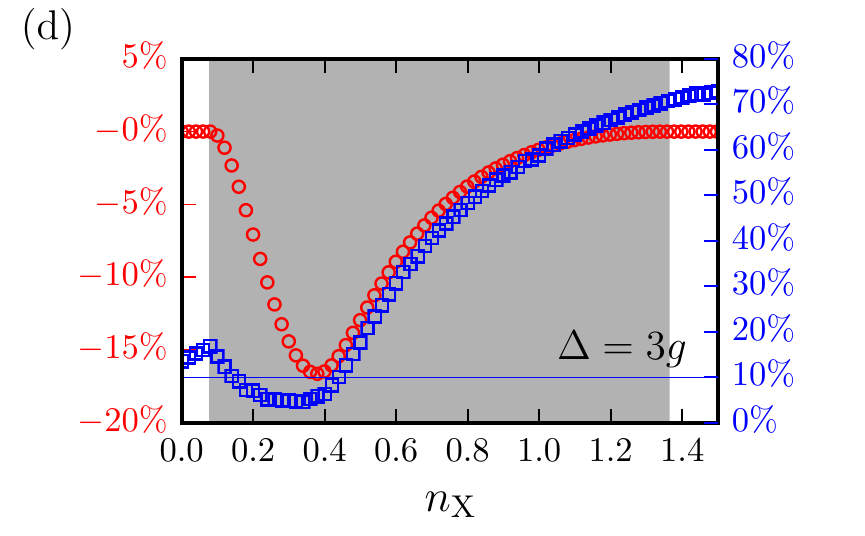}
\end{minipage}
\caption{{\bf Energetics and photon densities in the modulated phase.} Data denoted by red circles represent the quantity $r_{\rm e}$ in Eq.~(\ref{eq:delta}) plotted as a function of $n_{\rm X}$. Data denoted by blue squares represent the ratio between 
the photon density $n_{\rm ph}$ and the density $n_{\rm X}$ of total excitations, as a function of $n_{\rm X}$ and for the modulated phase only.
Data in this figure have been obtained by setting the dielectric constant at $\kappa_{\rm r}=5$ and the highest occupied LL at $M=1$. Above the horizontal blue line, $n_{\rm ph}/n_{\rm X}> 10\%$. Different panels refer to different values of the ratio $\Delta/g$: panel ({\bf a}) $\Delta/g = 2$, ({\bf b}) $2.5$, ({\bf c}) $2.75$, and ({\bf d}) $3$.\label{fig:nph}}
\end{figure}

For this form of variational wave function we have fixed $\theta$, $\phi$, $\alpha$, $u$, $v$, $Q^\ast$, and $\varphi$ by minimizing ${\cal E}$.  A summary of our main numerical results for $u$, $v$, $\theta$, and $Q^\ast$ is reported in Fig.~\ref{fig:jn_kr5} for two values of the detuning $\Delta$.
Minimization yields $\varphi=0$, $\phi = -\pi/2$, and $\arg(\alpha)=\pi-\phi$.  The dependence of $|\alpha|$ on $n_{\rm X}$ is given by $|\alpha|=\sqrt{n_{\rm X} - \left[1-\cos(\theta) J_0(u) \right]/2}$, where $J_0(x)$ is the Bessel function of order zero.
Figs.~\ref{fig:jn_kr5}({\bf a}) and~({\bf b}) illustrate the weak dependence of the characteristic wave number $Q^\ast$ on the density $n_{\rm X}$ of total excitations. In Fig.~\ref{fig:nph} we report, for each value of $n_{\rm X}$, 
the ratio
\begin{equation}\label{eq:delta}
r_{\rm e} = \frac{\epsilon_{\rm m} - \epsilon_{\rm h}}{|\epsilon_{\rm h} - \epsilon_{0}|}~.
\end{equation}
The numerator in Eq.~(\ref{eq:delta}) is the difference between the energy of the condensed modulated phase, $\epsilon_{\rm m}$, described by Eq.~(\ref{eq:Qmod}), and that of the condensed homogeneous phase, $\epsilon_{\rm h}$, described by Eqs.~(\ref{eq:argalpha})-(\ref{eq:sint}). In the condensed modulated phase, it follows that $r_{\rm e}<0$. The denominator in Eq.~(\ref{eq:delta}) is the condensate energy in the homogeneous phase. In Fig.~\ref{fig:nph} we clearly see that, depending on the detuning $\Delta$ and the density of excitations $n_{\rm X}$, the modulated phase comes with a condensate energy gain $r_{\rm e}$ in the window $\approx 5\%$-$15\%$, with values of the photon fraction that are well above $10\%$.

{\bf Discussion.} In this Article we have made two major simplifying approximations that deserve a detailed discussion. We have i) used a quasi-equilibrium approach based on a grand-canonical Hamiltonian and ii) treated electron-electron interactions at the mean-field level. 

i) Exciton-polariton condensates differ from ultracold atomic gases
in that the condensing quasiparticles have relatively short lifetimes, mainly because of photon losses in the cavity or metamaterial. 
External optical pumping is therefore needed to maintain a non-equilibrium steady state. It has been shown~\cite{keeling_bookchapter_2010} that the  
resulting non-equilibrium steady state can be approximated by a thermal equilibrium state 
when the thermalization time is shorter than the exciton polariton lifetime. 
Equilibrium approximations have been successfully used in the literature to describe exciton-polariton fluids in 
semiconductor microcavities~\cite{carusotto_rmp_2013,su_prl_2014,eastham_prb_2001,keeling_prl_2004,keeling_prb_2005}. 
Experimental studies in GaAs quantum wells have shown that the thermalization time 
criterion is satisfied above a critical pump level~\cite{deng_prl_2006} and that polariton-polariton interactions (which are responsible for thermalization) 
are strong~\cite{sun_arxiv_2015}. We assume below
that a similar thermal equilibrium state can be achieved in graphene QHPFs. 
Because polaritons interact more strongly when they have a larger excitation fraction, quasi-equilibrium polariton condensates are expected to be more accessible experimentally when the cavity photon energy is higher than the bare exciton energy, i.e.~at positive detuning.

ii) The possibility of non-mean-field behavior in the matter degrees of freedom is an issue. 
Mean-field theory is accurate for dilute excitons at low temperatures~\cite{BEC_book}, but could fail at high exciton densities. In particular, the modulated phase we have found may undergo quantum melting. However, matter degrees of freedom at integer filling factors in the quantum Hall regime tend to be often well described by mean-field theory~\cite{Giuliani_and_Vignale}. The accuracy of mean-field theory is generally related to the restricted Hilbert space of LLs, which preclude the formation of competing correlated states with larger quantum fluctuations. There are several examples of interesting broken symmetry states in both semiconductor quantum wells and graphene that are accurately described by mean-field theory, including spin-polarized ferromagnetic states at odd filling factors~\cite{giuliani_prb_1985}, coherent quantum Hall bilayers in semiconductors systems with coupled quantum wells~\cite{eisenstein_nature_2004}, and spin-density wave states in neutral graphene~\cite{goerbig_rmp_2011}.  In 
some 
cases, the state selected by mean-field-theory energy minimization is the only state 
in the quantum Hall Hilbert space with a given set of quantum numbers, and therefore is exact. 
The situation here is similar to the coherent bilayer state~\cite{eisenstein_nature_2004} in that we have coherence between adjacent LLs. 

Finally, we mention that physics similar to that described in this Article is not expected to be limited to graphene but should equally occur in 2D electron gases in semiconductor (e.g.~GaAs) quantum wells. There are a number of quantitative differences in detail, however.  Most critically, the anharmonic LL spectrum of graphene should make it possible to achieve a better selective coupling to a particular $M, M+1$ doublet~\cite{pellegrino_prb_2014}.

{\bf Data availability.} The data files used to prepare the figures shown in the manuscript are available from the corresponding author upon request..

\vspace*{5mm}
\textbf{Acknowledgments.}
This work was supported by ``Centro Siciliano di Fisica Nucleare e Struttura della Materia'' (CSFNSM), 
a 2012 Scuola Normale Superiore Internal Project, Fondazione Istituto Italiano di Tecnologia, and the European Union's Horizon 2020 research and innovation programme under Grant Agreement No. 696656 ``GrapheneCore1''. Work in Austin was supported by the DOE Division of Materials Sciences and Engineering under grant DE-FG03- 02ER45958, and by the Welch foundation under grant TBF1473. It is a great pleasure to acknowledge Rosario Fazio for early contributions to this work.
\vspace*{5mm}
\vspace*{5mm}
\textbf{Author contributions.}
All the authors conceived the work, agreed on the approach to pursue, analysed and discussed the results; 
F.M.D.P. performed the analytical and numerical calculations;  V.G., A.H.M., and M.P. supervised the work.
\vspace*{5mm}

%


\newpage


\clearpage


\section*{Supplementary material}

# Supplementary Figures

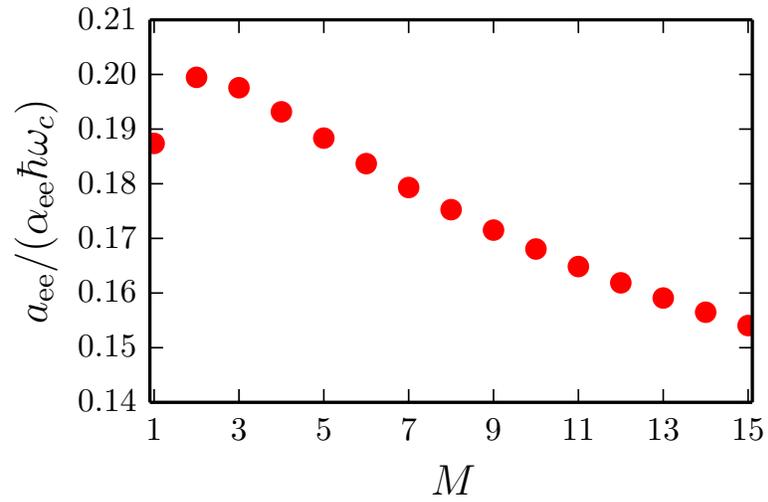

Supplementary Figure 1: **Dimensionless strength of electron-electron interactions.** Dependence of $a_{ee}$ (in units of $\alpha_{ee}\hbar\omega_c$) on the Landau level index $M$.

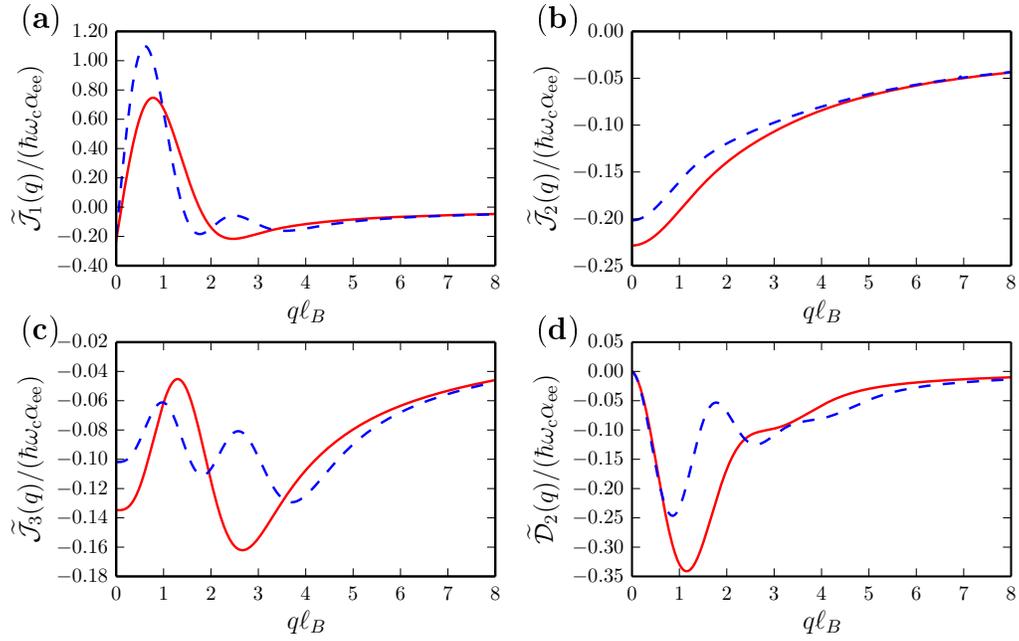

Supplementary Figure_2: **Pseudospin-pseudospin interactions.** Fourier transforms $\widetilde{\mathcal{J}}_\ell(q)$ and $\widetilde{\mathcal{D}}_2(q)$ (in units of $\hbar\omega_{\mathrm{c}}\alpha_{\mathrm{ee}}$). Panels (**a**), (**b**) and (**c**) refer to the symmetric interactions, while panel (**d**) refers to the antisymmetric interaction. The red solid (blue dashed) line refers to the case with highest occupied LL $M = 1$ ($M = 2$) in conduction band $\lambda = +$.

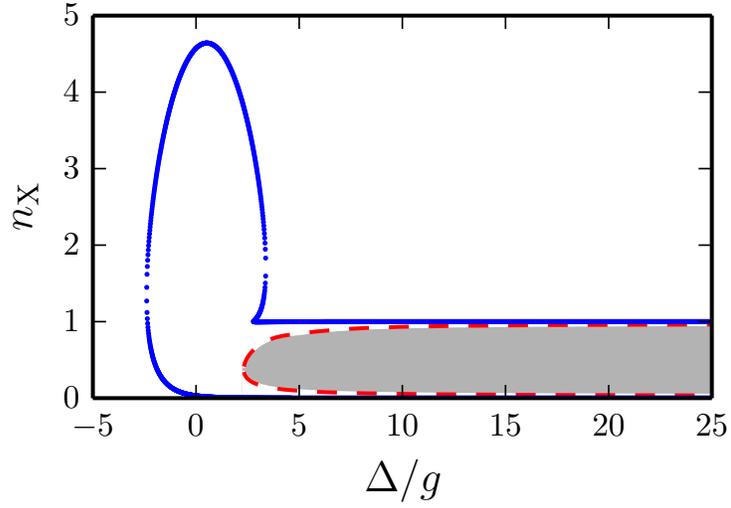

Supplementary Figure 3: **Phase diagram in the absence of symmetric interactions.** The red dashed boundary of the grey-shaded area—which represents the unstable region for $\mathcal{J}_\ell \neq 0$ as in Figure 2(**c**) of the main text—enlarges: for $\mathcal{J}_\ell = 0$ (absence of symmetric interactions) the new boundary is represented by the blue filled circles.

3

# Supplementary Note 1

At low energies, charge carriers in graphene are modeled by the usual single-channel massless Dirac fermion (MDF) Hamiltonian [1, 2]

$$\mathcal{H}_D = v_D \sigma \cdot \mathbf{p} \ , \tag{1}$$

where $v_D \approx 10^6$ m/s is the Dirac velocity. Here $\sigma = (\sigma_x, \sigma_y)$ is a 2D vector of Pauli matrices acting on sublattice degrees-of-freedom and $\mathbf{p} = -i\hbar\nabla_\mathbf{r}$ is the 2D momentum measured from one of the two corners (valleys) of the Brillouin zone.

A quantizing magnetic field $\mathbf{B} = B\hat{\mathbf{z}}$ perpendicular to the graphene sheet is coupled to the electronic degrees-of-freedom by replacing the canonical momentum $\mathbf{p}$ in Eq. (1) with the kinetic momentum $\mathbf{\Pi} = \mathbf{p} + e\mathbf{A}_0/c$, where $\mathbf{A}_0$ is the vector potential that describes the static magnetic field $\mathbf{B}$. The corresponding Hamiltonian is

$$\mathcal{H}_0 = v_D \sigma \cdot \mathbf{\Pi} \ . \tag{2}$$

We work in the Landau gauge $\mathbf{A}_0 = -By\hat{\mathbf{x}}$. In this gauge the canonical momentum along the $\hat{\mathbf{x}}$ direction, $p_x$, coincides with the magnetic translation operator [3] along the same direction and it commutes with the Hamiltonian $\mathcal{H}_0$. Thus, the eigenvalues of $p_x$ are good quantum numbers. A complete set of eigenfunctions of the Hamiltonian $\mathcal{H}_0$ in Eq. (2) is provided by the two component pseudospinors [4]

$$\langle \mathbf{r}|\lambda, n, k\rangle = \frac{e^{ikx}}{\sqrt{2L}} \left( \begin{array}{c} w_{-,n}\phi_{n-1}(y - \ell_B^2 k) \\ \lambda w_{+,n}\phi_n(y - \ell_B^2 k) \end{array} \right) \ , \tag{3}$$

where $\lambda = + \ (-)$ denotes conduction (valence) band levels, $n \in \mathbb{N}$ is the Landau level (LL) index, and $k$ is the eigenvalue of the magnetic translation operator in the $\hat{\mathbf{x}}$ direction. In Eq. (3)

$$w_{\pm,n} = \sqrt{1 \pm \delta_{n,0}} \tag{4}$$

guarantees that the pseudospinor corresponding to the $n = 0$ LL has nonzero weight only on one sublattice. Furthermore, $\phi_n(y)$ with $n = 0, 1, 2, \ldots$ are the normalized eigenfunctions of a 1D harmonic oscillator with frequency equal to the MDF cyclotron frequency $\omega_c = \sqrt{2}v_D/\ell_B$. Here $\ell_B = \sqrt{\hbar c/(eB)} \simeq 25$ nm$/\sqrt{B[\text{Tesla}]}$ is the magnetic length.

The spectrum of the Hamiltonian (2) has the well-known form [4]

$$\varepsilon_{\lambda,n} = \lambda \hbar \omega_c \sqrt{n} \ .$$ (5)

Each LL has a macroscopic degeneracy $\mathcal{N} = N_f S/(2\pi \ell_B^2) \equiv N_f \mathcal{N}_\phi$, where $N_f = 4$ is the spin-valley degeneracy and $S = L^2$ is the sample area.

The fully microscopic matter Hamiltonian is written as

$$\mathcal{H}_{\mathrm{mat}} = \mathcal{H}_0 + \mathcal{H}_{\mathrm{ee}} \ .$$ (6)

Here, $\mathcal{H}_0$ is the second-quantized version of Eq. (2),

$$\mathcal{H}_0 = \sum_{\lambda,n,k,\xi} \varepsilon_{\lambda,n} c^\dagger_{\lambda,n,k,\xi} c_{\lambda,n,k,\xi} \ ,$$ (7)

where $c^\dagger_{\lambda,n,k,\xi}$ $(c_{\lambda,n,k,\xi})$ is a fermionic creation (annihilation) operator for an electron with band index $\lambda$, LL quantum number $n$, and eigenvalue of the magnetic translation operator along the $\hat{\mathbf{x}}$ direction equal to $k$. The collective index $\xi$ refers to the valley $(K, K')$ index and spin-projection along the $\hat{\mathbf{z}}$ direction. The second term in Eq. (6), $\mathcal{H}_{\mathrm{ee}}$, represents Coulomb interactions. This term can be written as

$$\mathcal{H}_{\mathrm{ee}} = \frac{1}{2L^2} \sum_{\mathbf{q}} v_{\mathbf{q}} \left[ n_{-\mathbf{q}} n_{\mathbf{q}} - n_{\mathbf{q}=\mathbf{0}} \right] \ ,$$ (8)

where $n_{\mathbf{q}}$ is the Fourier transform of the electronic density operator

$$n_{\mathbf{q}} \equiv \int d\mathbf{r} e^{-i\mathbf{q}\cdot\mathbf{r}} \psi^\dagger(\mathbf{r}) \psi(\mathbf{r}) \ ,$$ (9)

and $v_{\mathbf{q}}$ is the 2D Fourier transform of the Coulomb potential

$$v_{\mathbf{q}} = \frac{2\pi e^2}{\epsilon q} \ .$$ (10)

Here $\kappa_{\mathbf{r}}$ is the cavity dielectric constant. The real-space field operators $\psi^\dagger(\mathbf{r})$ and $\psi(\mathbf{r})$ have been introduced in the main text.

Selecting only terms that respect the SU(4) spin-valley symmetry, the e-e interaction term in Eq. (8) can be written in the following manner [4]

$$\begin{aligned}
\mathcal{H}_{\mathrm{ee}} &= \frac{1}{2} \frac{1}{L^2} \sum_{\mathbf{q}} \sum_{\lambda_i, n_i} \sum_{\xi,\xi',k,k'} e^{i\ell_B^2 q_y(k-k')} \mathcal{V}_{(\lambda_1 n_1),(\lambda_2 n_2),(\lambda_3 n_3),(\lambda_4 n_4)}(\mathbf{q}) \\
&\times c^\dagger_{\lambda_1,n_1,\xi,k} c^\dagger_{\lambda_2,n_2,\xi',k'-q_x} c_{\lambda_3,n_3,\xi',k'} c_{\lambda_4,n_4,\xi,k-q_x} \ ,
\end{aligned}$$ (11)

where

$$\mathcal{V}_{(\lambda_1,n_1),(\lambda_2,n_2),(\lambda_3,n_3),(\lambda_4,n_4)}(\mathbf{q}) = v_\mathbf{q} \mathcal{F}_{(\lambda_1,n_1),(\lambda_4,n_4)}(-\mathbf{q}) \mathcal{F}_{(\lambda_2,n_2),(\lambda_3,n_3)}(\mathbf{q}) \ . \quad (12)$$

The form factors $\mathcal{F}_{(\lambda,n),(\lambda',n')}(\mathbf{q})$ are given by [4]

$$\begin{aligned}
\mathcal{F}_{(\lambda,n),(\lambda',n')}(\mathbf{q}) &\equiv \frac{1}{2} \Bigg[ w_{-,n} w_{-,n'} D_{n-1,n'-1}\left(-\frac{\ell_B \bar{q}^*}{\sqrt{2}}\right) \\
&+ \lambda \lambda' w_{+,n} w_{+,n'} D_{n,n'}\left(-\frac{\ell_B \bar{q}^*}{\sqrt{2}}\right) \Bigg] \ .
\end{aligned} \quad (13)$$

Here, $\bar{q} = q_x + i q_y$ and $\bar{q}^* = q_x - i q_y$,

$$D_{n,n'}(z) \equiv \begin{cases} \sqrt{\dfrac{n'!}{n!}} z^{n-n'} e^{-|z|^2/2} L_{n'}^{(n-n')}(|z|^2), & \text{for } n \geq n' \\ D_{n',n}(-z^*), & \text{for } n < n' \end{cases} \ , \quad (14)$$

and $L_n^{(n-n')}(x)$ are generalized Laguerre polynomials [5].

We consider the integer quantum Hall regime in which a given number of LLs are fully occupied and the remaining ones are empty. Since the MDF Hamiltonian is particle-hole symmetric, we can consider, without loss of generality, the situation in which graphene is $n$-doped and the Fermi energy lies in the conduction band ($\lambda = +$). We denote by $n = M$ the highest occupied LL, and the lowest empty LL is $n = M + 1$ We are interested in the case in which cavity photons with energy $\hbar\omega$ are nearly resonant with the energy difference between the two conduction-band LLs $n = M, M+1$. In this limit, the fermionic Hilbert space can be reduced to the resonant doublet. From here on, we denote by $\tilde{c}_{\lambda,n,k,\xi}$ and $\tilde{c}^\dagger_{\lambda,n,k,\xi}$ operators for states which do not belong to the resonant doublet $M, M + 1$. We will keep using $c_{\lambda,n,k,\xi}$ and $c^\dagger_{\lambda,n,k,\xi}$ only for states which belong to the resonant doublet. We can rewrite the full microscopic Hamiltonian in Eq. (6) as the sum of three terms:

$$\mathcal{H}_{\mathrm{mat}} = \mathcal{H}_{\mathrm{d}} + \mathcal{H}_{\mathrm{m}} + \mathcal{H}_{\mathrm{dm}} \ . \quad (15)$$

The first term $\mathcal{H}_{\mathrm{d}}$ contains only fermionic field operators related to the resonant doublet $M, M+1$. The second term, $\mathcal{H}_{\mathrm{m}}$, contains only field operators of the type $\tilde{c}_{\lambda,n,k,\xi}, \tilde{c}^\dagger_{\lambda,n,k,\xi}$: these degrees of freedom play the role of a "medium" for the resonant doublet. The third term, $\mathcal{H}_{\mathrm{dm}}$, describes coupling between the resonant doublet and the medium degrees of freedom.

To obtain an *effective* matter Hamiltonian, we start from the fully microscopic Hamiltonian $\mathcal{H}_{\mathrm{mat}}$ and we treat in an exact fashion all the terms that involve field operators ($c_{\lambda,n,k,\xi}$ and $c^\dagger_{\lambda,n,k,\xi}$) acting only on the doublet $n = M, M+1$ in conduction band $\lambda = +$. All the other terms (containing $\tilde{c}_{\lambda,n,k,\xi}$ and $\tilde{c}^\dagger_{\lambda,n,k,\xi}$) are treated within the Hartree-Fock approximation [3]. The medium Hamiltonian is discarded. In the coupling term $\mathcal{H}_{\mathrm{dm}}$, we only keep terms that separately conserve the number of particles in the $M, M+1$ doublet and in the medium. These are terms of the form $\tilde{c}^\dagger c^\dagger c \tilde{c}$. Terms of the form $\tilde{c}^\dagger \tilde{c}^\dagger cc$ are discarded. We therefore replace

$$\tilde{c}^\dagger_{\lambda_1,n_1,k_1,\xi_1} c^\dagger_{+,n_2,k_2,\xi_2} c_{+,n_3,k_3,\xi_3} \tilde{c}_{\lambda_4,n_4,k_4,\xi_4} \rightarrow$$
$$\delta_{\lambda_1,\lambda_4} \delta_{n_1,n_4} \delta_{k_1,k_4} \delta_{\xi_1,\xi_4} \Theta(\epsilon_{+,M} - \epsilon_{\lambda_1,n_1}) c^\dagger_{+,n_2,k_2,\xi_2} c_{+,n_3,k_3,\xi_3} \ . \quad (16)$$

After straightforward algebraic manipulations we reach the final result for the effective matter Hamiltonian, which is best expressed in a pseudospin representation in which pseudospin "up" ("down") corresponds to the $M+1$ ($M$) LL. To this end, we define the following pseudospin operators:

$$\bar{\rho}^{\xi\xi'}_m(\mathbf{q}) = \sqrt{\frac{1}{\mathcal{N}_\phi}} \sum_{n,n'=M,M+1} \sum_k [\tau_m]_{nn'} c^\dagger_{+,n,k,\xi} c_{+,n',k+q_x,\xi'} \ e^{-i\ell_B^2 q_y(k+q_x/2)} \ , \quad (17)$$

and

$$S_m(\mathbf{q}) = \sqrt{\frac{1}{N_{\mathrm{f}}}} \sum_\xi \bar{\rho}^{\xi\xi}_m(\mathbf{q}) \ , \quad (18)$$

where $m = 0, \ldots, 3$ and $[\tau_m]_{nn'}$ labels the matrix elements of a four-vector $\tau_m$ of $2 \times 2$ matrices acting on the $M, M+1$ doublet, specifically $\tau_0$ represents the $2 \times 2$ identity matrix and $\tau_1, \tau_2, \tau_3$ represent the ordinary $2 \times 2$ Pauli matrices.

The final effective matter Hamiltonian reads as following:

$$\mathcal{H}_{\mathrm{mat}} = E_M^\star \sqrt{\mathcal{N}} S_0(0) + \frac{\Omega_M^\star}{2} \sqrt{\mathcal{N}} S_3(0) + \frac{1}{2} \sum_{m,m',\mathbf{q}} V_{mm'}(\mathbf{q}) S_m(-\mathbf{q}) S_{m'}(\mathbf{q}) \ . \quad (19)$$

Here, $E_M^\star \equiv E_M + \tilde{E}_M$ and $\Omega_M^\star \equiv \Omega_M + \tilde{\Omega}_M$ with $E_M = \hbar\omega_{\mathrm{c}}(\sqrt{M+1} + \sqrt{M})/2$. The terms $\tilde{E}_M$ and $\tilde{\Omega}_M$ are due to exchange interactions between

the resonant doublet $M, M+1$ and *occupied* LLs outside the doublet, i.e.

$$
\begin{aligned}
\tilde{E}_M &= -\frac{1}{2L^2} \sum_{\mathbf{q}} \sideset{}{'}\sum_{\lambda,n} \Bigg[ \mathcal{V}_{(+,M+1),(\lambda,n),(+,M+1),(\lambda,n)}(\mathbf{q}) \\
&+ \mathcal{V}_{(+,M),(\lambda,n),(+,M),(\lambda,n)}(\mathbf{q}) \Bigg]
\end{aligned}
\tag{20}
$$

and

$$
\begin{aligned}
\tilde{\Omega}_M &= -\frac{1}{2L^2} \sum_{\mathbf{q}} \sideset{}{'}\sum_{\lambda,n} \Bigg[ \mathcal{V}_{(+,M+1),(\lambda,n),(+,M+1),(\lambda,n)}(\mathbf{q}) \\
&- \mathcal{V}_{(+,M),(\lambda,n),(+,M),(\lambda,n)}(\mathbf{q}) \Bigg] ,
\end{aligned}
\tag{21}
$$

where the summation over $\lambda, n$ runs only values such that $\epsilon_{\lambda,n} < \epsilon_{+1,M}$.

The last term in Eq. (19) describes e-e interactions within the resonant doublet $M, M+1$, i.e.

$$
V_{mm'}(\mathbf{q}) = \frac{1}{4} \sum_{n_i = M, M+1} [\tau_m]_{n_1,n_4} [\tau_{m'}]_{n_2,n_3} \mathcal{V}_{(+,n_1),(+,n_2),(+,n_3),(+,n_4)}(\mathbf{q}) .
\tag{22}
$$

This is a $4 \times 4$ Hermitian matrix, which can be decomposed into its real and imaginary parts: $V_{mm'}(\mathbf{q}) = \mathrm{Re}[V_{mm'}(\mathbf{q})] + i\mathrm{Im}[V_{mm'}(\mathbf{q})]$, where

$$
\begin{cases}
\mathrm{Re}[V_{mm'}(\mathbf{q})] = \mathrm{Re}[V_{m'm}(\mathbf{q})] \\
\mathrm{Im}[V_{mm'}(\mathbf{q})] = -\mathrm{Im}[V_{m'm}(\mathbf{q})]
\end{cases} .
\tag{23}
$$

It is possible to show that every quantity $V_{mm'}(\mathbf{q})$ is either purely real or purely imaginary: $V_{mm'}(\mathbf{q})$ is purely imaginary if $m = 1$ or $2$ ($m = 0$ or $3$) and simultaneously $m' = 0$ or $3$ ($m' = 1$ or $2$); for any other value of $m, m'$ $V_{mm'}(\mathbf{q})$ is purely real.

# Supplementary Note 2

We consider a graphene sheet coupled to the electromagnetic field in a cavity. The Hamiltonian that describes the coupling between electrons and cavity

photons is

$$
\begin{aligned}
\mathcal{H}_{\text{int}} &= \frac{1}{\sqrt{\mathcal{N}}} \sum_{\lambda,\lambda',n,n',\xi,\xi',k,\mathbf{q},\nu} e^{i\ell_B^2 q_y(k+q_x/2)} g_{\mathbf{q}} \big[ \lambda w_{\lambda n} e^-_{\text{em}}(\mathbf{q},\nu) \delta_{n',n+1} \\
&+ \lambda' w_{\lambda' n'} e^+_{\text{em}}(\mathbf{q},\nu) \delta_{n',n-1} \big] \big( a_{\mathbf{q},\nu} + a^\dagger_{-\mathbf{q},\nu} \big) c^\dagger_{\lambda' n' k + q_x} c_{\lambda n k} \ , \quad (24)
\end{aligned}
$$

where $g_{\mathbf{q}} = \hbar\omega_c \sqrt{e^2/(2\epsilon L_z \hbar\omega_{\mathbf{q}})}$ is the light-matter interaction parameter, $e^\pm_{\text{em}}(\mathbf{q},\nu) = (\hat{\mathbf{x}} \pm i\hat{\mathbf{y}}) \cdot \mathbf{e}_{\text{em}}(\mathbf{q},\nu)$ where $\mathbf{e}_{\text{em}}(\mathbf{q},\nu)$ is a unit vector describing the linear polarization $\nu$ of the electromagnetic field, $\omega_{\mathbf{q}} = \sqrt{\omega^2 + c^2 q^2/\kappa_r}$ is the cavity photon dispersion relation, $\kappa_r$ is the cavity dielectric constant, $V = L_z L^2$ the volume of the cavity, and $L_z \ll L$ is the cavity length along the $\hat{\mathbf{z}}$ direction.

Among all the processes described by $\mathcal{H}_{\text{int}}$, we take into account only resonant terms which describe the photon-induced electronic transitions between LLs $M$ and $M+1$ in conduction band. This approximation is called rotating wave approximation (RWA). In the RWA, the light-matter interaction Hamiltonian becomes

$$
\mathcal{H}_{\text{int}} = \sqrt{2} \sum_{\mathbf{q}} g_{\mathbf{q}} \left[ a^\dagger_{\mathbf{q},\text{L}} S_-(\mathbf{q}) + a_{-\mathbf{q},\text{L}} S_+(\mathbf{q}) \right] \ , \quad (25)
$$

where $a_{\mathbf{q},\text{L}} = (a_{\mathbf{q},x} - i a_{\mathbf{q},y})/\sqrt{2}$ $[a^\dagger_{\mathbf{q},\text{L}} = (a^\dagger_{\mathbf{q},x} + i a^\dagger_{\mathbf{q},y})/\sqrt{2}]$ is the annihilation [creation] operator for a left-handed photon and $S_\pm(\mathbf{q}) \equiv [S_1(\mathbf{q}) \pm i S_2(\mathbf{q})]/2$. The term containing $S_+$ ($S_-$) describes transitions from LL $M$ ($M+1$) to LL $M+1$ ($M$) assisted by the annihilation (creation) of a left-handed photon. For $\mathbf{B} = B\hat{\mathbf{z}}$, the RWA selects the left circular polarization, $\mathbf{e}_{\text{em}}(\mathbf{q},\text{L}) = (\hat{\mathbf{x}} - i\hat{\mathbf{y}})/\sqrt{2}$: of course, the RWA will select the right circular polarization, $\mathbf{e}_{\text{em}}(\mathbf{q},\text{R}) = (\hat{\mathbf{x}} + i\hat{\mathbf{y}})/\sqrt{2}$, for $\mathbf{B} = -B\hat{\mathbf{z}}$.

Within the RWA, the sum of the number of cavity photons $N_{\text{ph}}$ and the number of excitons $N_{\text{ex}}$ is a *conserved* quantity. The number of photons can be expressed in terms of the photon field operator as

$$
N_{\text{ph}} \equiv \sum_{\mathbf{q},\nu} a^\dagger_{\mathbf{q},\nu} a_{\mathbf{q},\nu} \ , \quad (26)
$$

while the number the number of excitons can be expressed in terms of the pseudospin operators as

$$
N_{\text{ex}} \equiv \mathcal{N}[1 + S_3(0)/\sqrt{\mathcal{N}}]/2 \ . \quad (27)
$$

# Supplementary Note 3

It is possible to show that the variational state $|\psi\rangle$ introduced in Eq. (3) of the main text can be re-written as:

$$|\psi\rangle = e^{-\mathcal{N}\frac{|\gamma|^2}{2}} e^{\sqrt{\mathcal{N}}\gamma p^\dagger} |\psi_0\rangle \ , \tag{28}$$

where

$$p^\dagger \equiv \frac{1}{\gamma} \left( \alpha a^\dagger_{\mathbf{0},\text{L}} + \beta d^\dagger \right) \ , \tag{29}$$

$$d^\dagger \equiv \frac{1}{\sqrt{\mathcal{N}}\beta} \sum_{k,\xi} e^{-i\phi_k} \tan(\theta_k/2) c^\dagger_{+,M+1,k,\xi} c_{+,M,k,\xi} \ , \tag{30}$$

$$\beta \equiv \sqrt{-\frac{2}{\mathcal{N}} \sum_{k,\xi} \log(\cos\theta_k/2)} \ , \tag{31}$$

and

$$\gamma \equiv \sqrt{|\alpha|^2 + |\beta|^2} \ . \tag{32}$$

We note that the commutator between the operator $d^\dagger$ and its hermitian conjugate $d$ is given by:

$$[d, d^\dagger] = \frac{1}{\mathcal{N}\beta^2} \sum_{k,\xi} \tan^2(\theta_k/2) \left( c^\dagger_{+1,M,k,\xi} c_{+1,M,k,\xi} - c^\dagger_{+1,M+1,k,\xi} c_{+1,M+1,k,\xi} \right) \ . \tag{33}$$

Calculating the expectation value of $[d, d^\dagger]$ on the variational state introduced in Eq. (3) of the main text and taking the low-density limit $\theta_k \ll 1$, we find:

$$\langle\psi|[d, d^\dagger]|\psi\rangle \approx 1 - \frac{\sum_k \theta_k^4}{2\sum_k \theta_k^2} \ . \tag{34}$$

If the second term on the right-hand side of the previous equation is neglected, excitons can be treated as bosons. In the same limit, the operator $p^\dagger$ has the meaning of a polariton creation operator. The variational state reported in Eq. (3) of the main text is a coherent state of polaritons and the quantity $\mathcal{N}|\gamma|^2$ represents the average number of polaritons.

# Supplementary Note 4

We here report a number of relevant parameters that appear in the energy functional introduced in Eq. (7) of the main text:

$$
\begin{aligned}
\Delta_{\rm ee} & \equiv \frac{1}{L^2} \sum_{\mathbf{q}} v_{\mathbf{q}} \Bigg\{ \sum_{\lambda,n}{}' \left[ |\mathcal{F}_{(+,M),(\lambda,n)}(\mathbf{q})|^2 - |\mathcal{F}_{(+,M+1),(\lambda,n)}(\mathbf{q})|^2 \right] \\
& \quad - \ \mathcal{F}_{(+,M),(+,M)}(\mathbf{q}) \mathcal{F}_{(+,M+1),(+,M+1)}(\mathbf{q}) \Bigg\} \,,
\end{aligned}
\tag{35}
$$

$$
\begin{aligned}
a_{\rm ee} & \equiv \frac{1}{L^2} \sum_{\mathbf{q}} v_{\mathbf{q}} \Bigg\{ -\frac{1}{2} \left[ \mathcal{F}_{(+,M),(+,M)}(\mathbf{q}) - \mathcal{F}_{(+,M+1),(+,M+1)}(\mathbf{q}) \right]^2 \\
& \quad + \ |\mathcal{F}_{(+,M+1),(+,M)}(\mathbf{q})|^2 \Bigg\} \,,
\end{aligned}
\tag{36}
$$

and

$$
\begin{aligned}
\epsilon_0 & \equiv \frac{\Delta - a_{\rm ee}}{2} - \frac{1}{8L^2} \sum_{\mathbf{q}} v_{\mathbf{q}} \big[ 2 |\mathcal{F}_{(+,M+1),(+,M)}(\mathbf{q})|^2 \\
& \quad - \ |\mathcal{F}_{(+,M+1),(+,M+1)}(\mathbf{q})|^2 - |\mathcal{F}_{(+,M),(+,M)}(\mathbf{q})|^2 \big] \,.
\end{aligned}
\tag{37}
$$

As in Supplementary Note 1, the sum over $\lambda, n$ runs only over indices such that $\epsilon_{\lambda,n} < \epsilon_{+,M}$. We note that $a_{\rm ee}$ involves only the two resonant LLs $M$ and $M+1$.

Supplementary Figure 1 shows the quantity $a_{\rm ee}$ (in units of $\alpha_{\rm ee} \hbar \omega_{\rm c}$) as a function of the LL index $M$ in the interval $1 \leq M \leq 15$. Here $\alpha_{\rm ee} = e^2/(\kappa_{\rm r} \hbar v_{\rm D}) \approx 2.2/\kappa_{\rm r}$.

The symmetric and antisymmetric pseudospin-pseudospin interactions can be decomposed as following:

$$
\mathcal{J}_\ell(k - k') \equiv \mathcal{J}_{\ell,{\rm d}}(k - k') + \mathcal{J}_{\ell,{\rm x}}(k - k')
\tag{38}
$$

and

$$
\mathcal{D}_2(k - k') \equiv \mathcal{D}_{2,{\rm d}}(k - k') + \mathcal{D}_{2,{\rm x}}(k - k') \,.
\tag{39}
$$

The *direct* contributions are given by:

$$\mathcal{J}_{1,\mathrm{d}}(k-k') = \frac{\mathcal{N}}{L^2} \sum_q e^{i\ell_B^2 q(k-k')} v_q |\mathcal{F}_{(+,M+1),(+,M)}(q\hat{\boldsymbol{y}})|^2 \ , \tag{40a}$$

$$\mathcal{J}_{2,\mathrm{d}}(k-k') = 0 \ , \tag{40b}$$

$$\mathcal{J}_{3,\mathrm{d}}(k-k') = \frac{\mathcal{N}}{4L^2} \sum_q e^{i\ell_B^2 q(k-k')} v_q \big[\mathcal{F}_{(+,M+1),(+,M+1)}(q\hat{\boldsymbol{y}})$$

$$- \ \mathcal{F}_{(+,M),(+,M)}(q\hat{\boldsymbol{y}})\big]^2 \ , \tag{40c}$$

$$\mathcal{D}_{2,\mathrm{x}}(k-k') = \frac{\mathcal{N}}{2L^2} \sum_q e^{i\ell_B^2 q(k-k')} v_q \mathcal{F}_{(+,M+1),(+,M)}(q\hat{\boldsymbol{y}})$$

$$\times \ \big[\mathcal{F}_{(+,M+1),(+,M+1)}(q\hat{\boldsymbol{y}}) - \mathcal{F}_{(+,M),(+,M)}(q\hat{\boldsymbol{y}})\big] \ . \tag{40d}$$

The *exchange* contributions are given by:

$$\mathcal{J}_{1,\mathrm{x}}(k-k') = -\frac{\mathcal{N}_\phi}{2L^2} \sum_{\mathbf{q}} v_{\mathbf{q}} \big\{\mathcal{F}_{(+,M+1),(+,M+1)}(\mathbf{q})\mathcal{F}_{(+,M),(+,M)}(\mathbf{q})$$

$$- \ \mathrm{Re}^2\big[\mathcal{F}_{(+,M+1),(+,M)}(\mathbf{q})\big]\big\}\delta_{q_x,k-k'} \ , \tag{41a}$$

$$\mathcal{J}_{2,\mathrm{x}}(k-k') = -\frac{\mathcal{N}_\phi}{2L^2} \sum_{\mathbf{q}} v_{\mathbf{q}} \big\{\mathcal{F}_{(+,M+1),(+,M+1)}(\mathbf{q})\mathcal{F}_{(+,M),(+,M)}(\mathbf{q})$$

$$+ \ \mathrm{Re}^2\big[\mathcal{F}_{(+,M+1),(+,M)}(\mathbf{q})\big]\big\}\delta_{q_x,k-k'} \ , \tag{41b}$$

$$\mathcal{J}_{3,\mathrm{x}}(k-k') = \frac{\mathcal{N}_\phi}{4L^2} \sum_{\mathbf{q}} v_{\mathbf{q}} \big[2|\mathcal{F}_{(+,M+1),(+,M)}(\mathbf{q})|^2 - |\mathcal{F}_{(+,M+1),(+,M+1)}(\mathbf{q})|^2$$

$$- \ |\mathcal{F}_{(+,M),(+,M)}(\mathbf{q})|^2\big]\delta_{q_x,k-k'} \ , \tag{41c}$$

$$\mathcal{D}_{2,\mathrm{x}}(k-k') = -\frac{\mathcal{N}_\phi}{2L^2} \sum_{\mathbf{q}} v_{\mathbf{q}} \big[\mathcal{F}_{(+,M+1),(+,M+1)}(\mathbf{q}) + \mathcal{F}_{(+,M),(+,M)}(\mathbf{q})\big]$$

$$\times \ \mathrm{Re}\big[\mathcal{F}_{(+,M+1),(+,M)}(\mathbf{q})\big]\delta_{q_x,k-k'} \ , \tag{41d}$$

where $\mathcal{N}_\phi$ has been introduced in the main text.

Supplementary Figures 2 (**a**), (**b**), (**c**) and (**d**) show the Fourier transforms

$$\widetilde{\mathcal{J}}_\ell(q) \equiv \frac{1}{\mathcal{N}_\phi} \sum_k \mathcal{J}_\ell(k) e^{-iqk\ell_B^2} \tag{42}$$

and

$$\widetilde{\mathcal{D}}_2(q) \equiv \frac{1}{\mathcal{N}_\phi} \sum_k \mathcal{D}_2(k) e^{-iqk\ell_B^2} \ , \tag{43}$$

for $M = 1$ (red solid line) and $M = 2$ (blue dashed line). In the $q \to 0$ limit we obtain: $\widetilde{\mathcal{D}}_2(0) = 0$, $\widetilde{\mathcal{J}}_\ell(0) = \widetilde{\mathcal{J}}_{\ell,\mathrm{x}}(0) \neq 0$ for any $\ell$, and $\widetilde{\mathcal{J}}_1(0) = \widetilde{\mathcal{J}}_2(0)$. Moreover, we find that

$$a_{\mathrm{ee}} = 2\big[\widetilde{\mathcal{J}}_3(0) - \widetilde{\mathcal{J}}_1(0)\big] \ . \tag{44}$$

# Supplementary Note 5

The correction $\Delta_{\mathrm{ee}}$ in Eq. (35) to the cyclotron transition energy is related to the renormalization [6, 7, 8] of the Dirac velocity $v_{\mathrm{D}}$ due to exchange interactions, which occurs also in the absence of a magnetic field. It is well-known [9, 10, 11] that $\Delta_{\mathrm{ee}}$ is logarithmically divergent, i.e.

$$\Delta_{\mathrm{ee}} = \alpha_{\mathrm{ee}} \frac{\Omega_M}{8} \left[\ln(n_{\max}) + C_M\right] \ , \tag{45}$$

where $n_{\max}$ is a cut-off defined below and the constant $C_M$ depends on the highest-occupied LL with index $M$. For example, $C_0 \simeq -1.017$ and $C_1 \simeq -2.510$. The Dirac model applies over a large but finite energy region, so we define a high-energy cut-off $W$ in valence band. At any given magnetic field $B$, the integer $n_{\max}$ represents the number of LLs in the valence band with energy larger than $W$, i.e.

$$n_{\max} = \frac{B_W}{B} \ , \tag{46}$$

where $B_W = eW^2/(2c\hbar v_{\mathrm{D}})$, such that $W = \sqrt{2}\hbar v_{\mathrm{D}}\sqrt{n_{\max}}/\ell_B$. We follow Shizuya [11] in writing the correction to cyclotron transition energy in terms of the renormalized Dirac velocity: $\Omega_M = \sqrt{2}\hbar v_{\mathrm{D}}^\star/\ell_B$ and $v_{\mathrm{D}}^\star = v_{\mathrm{D}} + \delta v_{\mathrm{D}}$. The quantity $v_{\mathrm{D}}$ is the bare Dirac velocity and $\delta v_{\mathrm{D}} = \alpha_{\mathrm{QED}}c/(8\kappa_{\mathrm{r}})[\ln(B_W/B) + C_M]$ is the correction to the Dirac velocity, where $\alpha_{\mathrm{QED}} = e^2/(\hbar c) \simeq 1/137$ is the QED fine-structure constant. This allows us to fix $B_W = 450$ Tesla to make sure that $v_{\mathrm{D}}^\star$ matches the value $v_{\mathrm{D}}^\star = 1.12 \times 10^6$ m/s measured [12] from the intraband $0 \to 1$ cyclotron transition energy at $B = 18$ Tesla and in a sample with $\kappa_{\mathrm{r}} = 5$. We have fixed the bare Dirac velocity to $v_{\mathrm{D}} = c/300$.

# Supplementary Note 6

In this Supplementary Note we discuss the relative role of symmetric and antisymmetric pseudospin-pseudospin interactions in determining the precise form of the phase diagrams shown in Figure 2 of the main text.

13

We start by setting to zero all the symmetric interactions in the energy functional introduced in Eq. (7) of the main text: $\mathcal{J}_\ell = 0$. In this case, we find that the unstable regions (grey-shaded regions in Figure 2 of the main text) considerable expand. This is illustrated for the case $\kappa_{\mathrm{r}} = 15$ and $M = 1$ in Supplementary Figure 3.

## Supplementary Note 7

In this Supplementary Note we explain more in detail the approach we have followed to find the elementary excitations of the polariton fluid.

In the spin-chain language introduced in the main text, the state $|\psi\rangle$ of the homogeneous fluid phase represents a *collinear ferromagnet* in which the expectation value of the spin operator $\mathbf{S}(\mathbf{q}) \equiv [S_1(\mathbf{q}), S_2(\mathbf{q}), S_3(\mathbf{q})]^{\mathrm{T}}$ is non-zero only at $\mathbf{q} = \mathbf{0}$ and oriented along the direction

$$\hat{\mathbf{z}}' = [-\cos(\phi)\sin(\theta), -\sin(\phi)\sin(\theta), \cos(\theta)]^{\mathrm{T}} , \qquad (47)$$

i.e. $\langle\psi|\mathbf{S}(\mathbf{q})|\psi\rangle = -\hat{\mathbf{z}}'\sqrt{\mathcal{N}}\delta_{\mathbf{q},0}$. The collection of units vectors $\hat{\mathbf{x}}'$, $\hat{\mathbf{y}}'$, $\hat{\mathbf{z}}'$ with

$$\hat{\mathbf{x}}' = [\cos(\phi)\cos(\theta), \sin(\phi)\cos(\theta), \sin(\theta)]^{\mathrm{T}} , \qquad (48)$$

and

$$\hat{\mathbf{y}}' = [-\sin(\phi), \cos(\phi), 0]^{\mathrm{T}} , \qquad (49)$$

forms an orthonormal set, which will be used below to construct low-energy collective excitations above the ground state $|\psi\rangle$.

To study the collective mode spectrum, we assume that $|\psi\rangle$ is subject to an infinitesimal pertubation, which induces a small change $|\delta\psi\rangle$, i.e. we write $|\psi'\rangle = |\psi\rangle + |\delta\psi\rangle$. The most general infinitesimal change $|\delta\psi\rangle$ *orthogonal* to $|\psi\rangle$ can be written as a superposition of low-energy single-particle excitations [13]

$$|\delta\psi\rangle = \left(\sum_{\nu,\mathbf{q}} \alpha_{\mathbf{q}\nu} a^\dagger_{\mathbf{q},\nu} + \sum_{\xi,\mathbf{q}} r_{\mathbf{q}\xi} \bar{\rho}^{\xi\xi}_+(-\mathbf{q})\right)|\psi\rangle , \qquad (50)$$

where $\bar{\rho}^{\xi\xi}_+(\mathbf{q}) = \bar{\rho}^{\xi\xi}(\mathbf{q}) \cdot (\hat{\mathbf{x}}' + i\hat{\mathbf{y}}')/2$ and $\bar{\rho}^{\xi\xi}(\mathbf{q}) \equiv [\bar{\rho}^{\xi\xi}_1(\mathbf{q}), \bar{\rho}^{\xi\xi}_2(\mathbf{q}), \bar{\rho}^{\xi\xi}_3(\mathbf{q})]^{\mathrm{T}}$, with $\bar{\rho}^{\xi\xi}_m(\mathbf{q})$ as in Eq. (17). In writing Eq. (50) we have assumed that the infinitesimal pertubation does not induce any spin-flip or intervalley scattering process.

In the time-dependent Hartree-Fock (or generalized random phase) approximation [13], the spin wave spectrum can be calculated by making use of the Heisenberg equation of motion (EOM) for the expectation value of an operator $\mathcal{O}$, evaluated on the state $|\psi'\rangle$, i.e. $i\hbar\partial_t\langle\psi'|\mathcal{O}|\psi'\rangle = \langle\psi'|[\mathcal{O},\mathcal{H}]|\psi'\rangle$, where the total Hamiltonian $\mathcal{H}$ is reported in Eq. (1) of the main text. Writing EOMs for $\mathcal{O} = a_{\mathbf{q},\nu}$, $\bar{\rho}_-^{\xi\xi}(\mathbf{q})$, $a_{\mathbf{q},\nu}^\dagger$, and $\bar{\rho}_+^{\xi\xi}(\mathbf{q})$ and keeping terms up to linear order in $\alpha_{\mathbf{q}\nu} = \langle a_{\mathbf{q},\nu}\rangle$ and $r_{\mathbf{q}\xi} = \langle\bar{\rho}_-^{\xi\xi}(\mathbf{q})\rangle$, we find a homogeneous system of first-order linear differential equations. Eigenmodes are found by replacing $i\hbar\partial_t \to \Omega$. We obtain an eigenvalue problem, whose solution gives the spin wave spectrum. We find six independent collective modes, four due to the fourfold spin-valley degeneracy and the remaining two due to the two possible light polarizations. Only two collective modes are *hybrid* in that they contain both light and matter components. They are composed by a left-handed photon $a_{\mathbf{q},\mathrm{L}}$ and a bright collective matter mode $S_-'(\mathbf{q}) = \mathbf{S}(\mathbf{q})\cdot(\hat{\mathbf{x}}'-i\hat{\mathbf{y}}')/2$. In this subspace of hybrid modes, the eigenvalue problem reduces to $(\Omega_\mathbf{q}\mathbb{1} - \mathbf{M})\mathbf{v} = 0$ where $\mathbb{1}$ is the $4\times 4$ identity and the $4\times 4$ matrix $\mathbf{M}$ is reported in Eq. (22) of the main text.

# Supplementary References



\end{document}